# Super-adiabatic Temperature Gradient at Jupiter's Equatorial Zone and Implications for the Water Abundance

Cheng Li[1,*], Michael Allison[2], Sushil Atreya[1], Shawn Brueshaber[3], Leigh N. Fletcher[4], Tristan Guillot[5], Liming Li[6], Jonathan Lunine[7,8], Yamila Miguel[9,10], Glenn Orton[11], Paul Steffes[12], J. Hunter Waite[13], Michael H. Wong[14,15], Steven Levin[11], Scott Bolton[16]

[1]University of Michigan, Ann Arbor, USA

[2]Columbia University, New York, NY, USA

[3]Michigan Technical University, Houghton, MI, USA

[4]School of Physics and Astronomy, University of Leicester, University Road, Leicester, LE1 7RH, UK

[5]Observatoire de la Côte d'Azur, France

[6]University of Houston, USA

[7]Department of Astronomy, Cornell University, USA

[8]Carl Sagan Institute, Cornell University, USA

[9]Leiden Observatory, Leiden University, P.O. Box 9513, 2300 RA, Leiden, The Netherlands

[10]SRON Netherlands Institute for Space Research, Niels Bohrweg 4, 2333 CA Leiden, The Netherlands

[11]Jet Propulsion Laboratory, California Institute of Technology, USA

[12]Georgia Institute of Technology, USA

[13]Waite Science LLC, USA

[14]SETI Institute, USA

[15]University of California, Berkeley, USA

[16]Southwest Research Institute, USA

* Corresponding author: chengcli@umich.edu

**Abstract**

The temperature structure of a giant planet was traditionally thought to be an adiabat assuming convective mixing homogenizes entropy. The only in-situ measurement made by the Galileo Probe detected a near-adiabatic temperature structure within one of Jupiter's $5\mu m$ hot spots with small but definite local departures from adiabaticity. We analyze Juno's microwave observations near Jupiter's equator (0 ~ 5 °N) and find that the equatorial temperature structure is best characterized by a stable super-adiabatic temperature profile rather than an adiabatic one. Water is the only substance with sufficient abundance to alter the atmosphere's mean molecular weight and prevent dynamic instability if a super-adiabatic temperature gradient exists. Thus, from the super-adiabaticity, our results indicate a water concentration (or the oxygen to hydrogen ratio) of about 4.9 times solar with a possible range of 1.5 ~ 8.3 times solar in Jupiter's equatorial region.

# 1    Introduction

The Microwave Radiometer (MWR) onboard the Juno spacecraft has measured the brightness temperatures of Jupiter's thermal emission at six different microwave frequencies ranging from 0.6-22 GHz that cover a pressure range from ~250 bars to 0.5 bar (Janssen et al., 2017).One of the unique advantages of the MWR is its capability to measure the angular dependence of the thermal emission, known as limb darkening, which provides important additional information that is inaccessible from ground-based radio observations. We define limb darkening as the fractional reduction ($R$) of the microwave radiance (which we express as brightness temperature) from nadir viewing ($T_{b,nadir}$) to limb viewing ($T_{b,\theta}$) at emission angle $\theta$, i.e. $R_\theta = (T_{b,nadir} - T_{b,\theta})/T_{b,nadir}$. So, $R_{45}$ denotes the limb darkening evaluated at a 45-degree emission angle. Initial analyses of the MWR data have yielded the concentration of ammonia as a function of latitude and altitude using the nadir brightness temperature only and the abundance of water in Jupiter's equatorial zone (EZ) assuming a moist adiabatic temperature profile (Li et al., 2017, 2020), i.e. a temperature profile that is sub-adiabatic because of latent heat release during water condensation.

Although a moist adiabatic temperature profile was established by observing Earth's tropical climate (Xu & Emanuel, 1989), the Galileo Probe Atmospheric Structure Instrument (ASI) found significant departures from an ideal adiabat at Jupiter (Magalhães et al., 2002). Specifically, a sub-adiabatic temperature gradient was found between 0.5 to 1.7 bars, 3 to 8.5 bars and 14 to 20 bars with a static stability of 0.1~ 0.2 K/km. Given a scale height of ~30 km for Jupiter's atmosphere, the observed static stability at the Galileo Probe site would imply a departure from the dry adiabatic temperature profile by 3 ~ 6 K between 3 to 10 bars, and 10 to 20 K between 1 to 20 bars (3 scale heights). In contrast, a similar static stability on Earth would only lead to a temperature anomaly of around 1 K between the top and bottom of the troposphere due to a much thinner atmosphere on Earth.

Hence, the deviation from adiabaticity is potentially detectable by the Juno microwave radiometric observation because the instrument thermal noise is less than 1 K and an absolute calibration at pressures considered here (1~20 bars) is conservatively determined to be around 2% (Janssen et al., 2017). Figure 1 shows calculations of brightness temperatures and limb darkening at six Juno MWR channels for three idealized Jupiter's atmospheric models at the equator: a sub-adiabatic model, a dry adiabatic model and a super-adiabatic model. Each model assumes a homogeneous atmospheric composition in thermodynamic equilibrium without cloud condensates and the non-adiabatic part of the atmosphere is restricted to be between 1 to 20 bars at $\pm$ 0.15 K/km. At pressures higher than 20 bars, the atmospheric temperature is assumed to be dry adiabatic.

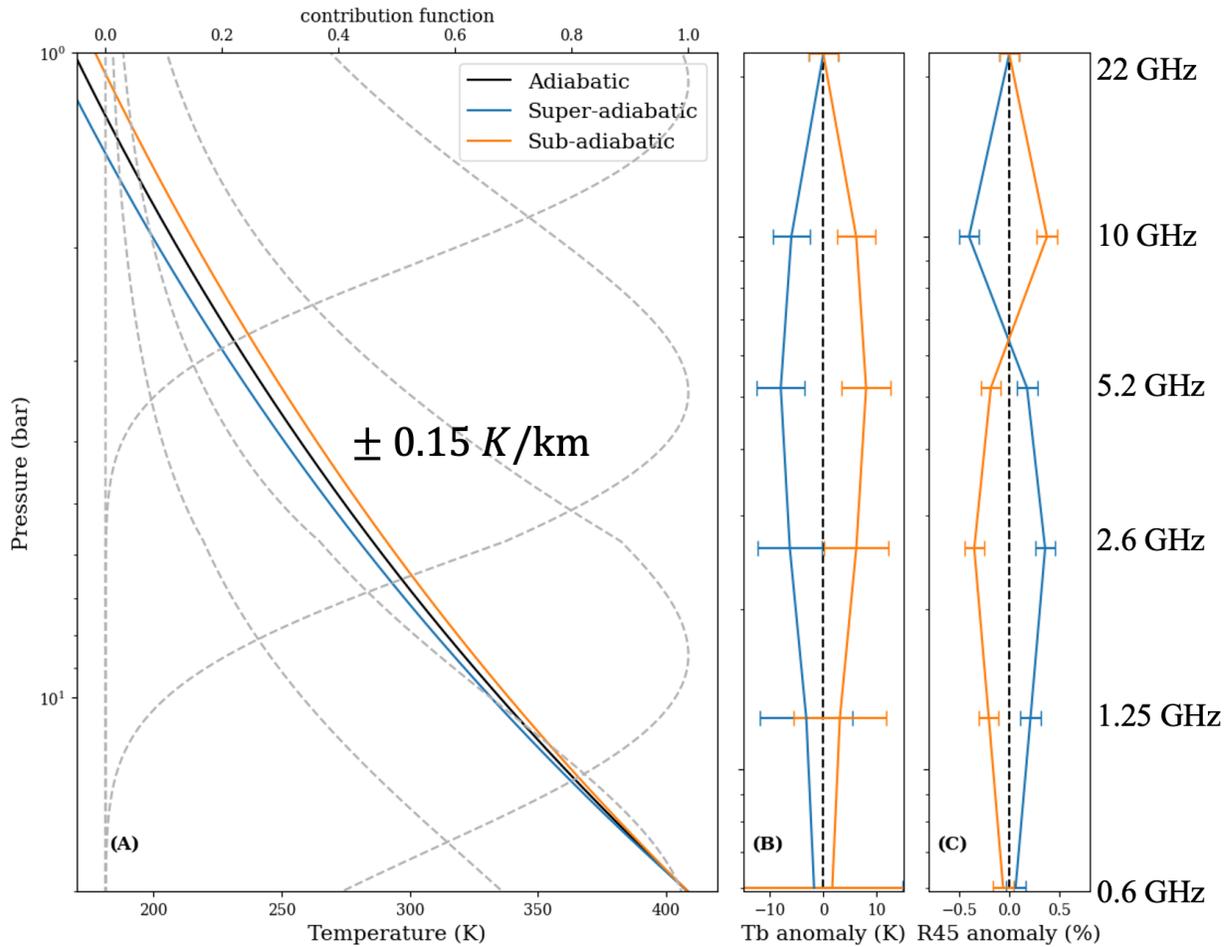

*Figure 1. (A) Three idealized Jovian thermal models, dry adiabatic (black), sub-adiabatic (orange) and super-adiabatic (blue). The non-adiabatic temperature gradient is $\pm$ 0.15 K/km. The grey dashed lines show the contribution functions of 10, 5.2, 2.6, and 1.25 GHz channels (from top to bottom). (B) Nadir brightness temperature anomalies at six Juno MWR channels. From top to bottom, they are at 22 GHz, 10 GHz, 5.2 GHz, 2.6 GHz, 1.25 GHz, and 0.6 GHz, respectively. The horizontal error bars show a conservative 2% uncertainty.* The contribution

functions for 22 GHz and 0.6 GHz peak at pressures outside the range of the graph. *(C) Limb-darkening anomalies (evaluated at a 45-degree emission angle). The horizontal error bars show a 0.1% instrument noise level (Janssen et al., 2005, 2017).*

These calculations show that brightness temperatures and limb darkening respond approximately linearly to the change in the temperature gradient such that a super-adiabatic model exhibits deviations in these parameters relative to an adiabat almost exactly the opposite of a sub-adiabatic model. The brightness temperatures in a super-adiabatic model are consistently colder across all frequencies compared to a dry adiabatic model, with the most significant difference observed for the 5.2-GHz channel. In contrast to the brightness temperature, the limb darkening anomalies change sign between the 10-GHz and the 2.6-GHz channels, because limb darkening is sensitive to the change of brightness temperature with pressure.

Brightness temperature is determined by the kinetic temperature of the atmosphere and the concentration of ammonia and water vapor. At higher altitudes ($p < 0.7$ bar), the concentration of ammonia vapor is limited by its saturation vapor pressure, dependent solely on temperature. Therefore, an increased temperature gradient across the ammonia condensation level suggests a larger ammonia gradient as well, which contributes mainly to the limb darkening anomaly observed at the 10-GHz channel. On the other hand, at lower altitudes ($p > 5$ bar), the concentration of ammonia is nearly constant in our model. The 2.6-GHz channel predominantly senses the kinetic temperature gradient, and its limb-darkening effect contrasts with the ammonia gradient sensed by the 10-GHz channel. At the 5.2-GHz channel, the above two effects approximately cancel one another, resulting in a minimum in the limb-darkening anomaly among these three channels.

We can perform a similar exercise for changing the vertical gradient of ammonia concentration. Still using the 20-bar pressure level as the base below which the atmosphere becomes homogenous, we try three simple models of ammonia profiles, homogenous (equilibrium condensation), positive gradient and negative gradient with height. Based on prior studies (Li et al., 2017, 2020), both positive and negative gradients have been identified in Jupiter's atmosphere between 1 and 20 bars at varying latitudes. Specifically, the equatorial region of Jupiter appears to exhibit a positive ammonia gradient, while other latitudes display a negative gradient. Examining the forward calculation results shown in Figure 2, the model with a negative ammonia gradient has a similar effect as a sub-adiabatic temperature profile in terms of brightness temperature and limb darkening observations. This demonstrates a degeneracy between the ammonia gradient and the temperature gradient that potentially limits what we can conclude from the spectral inversion study.

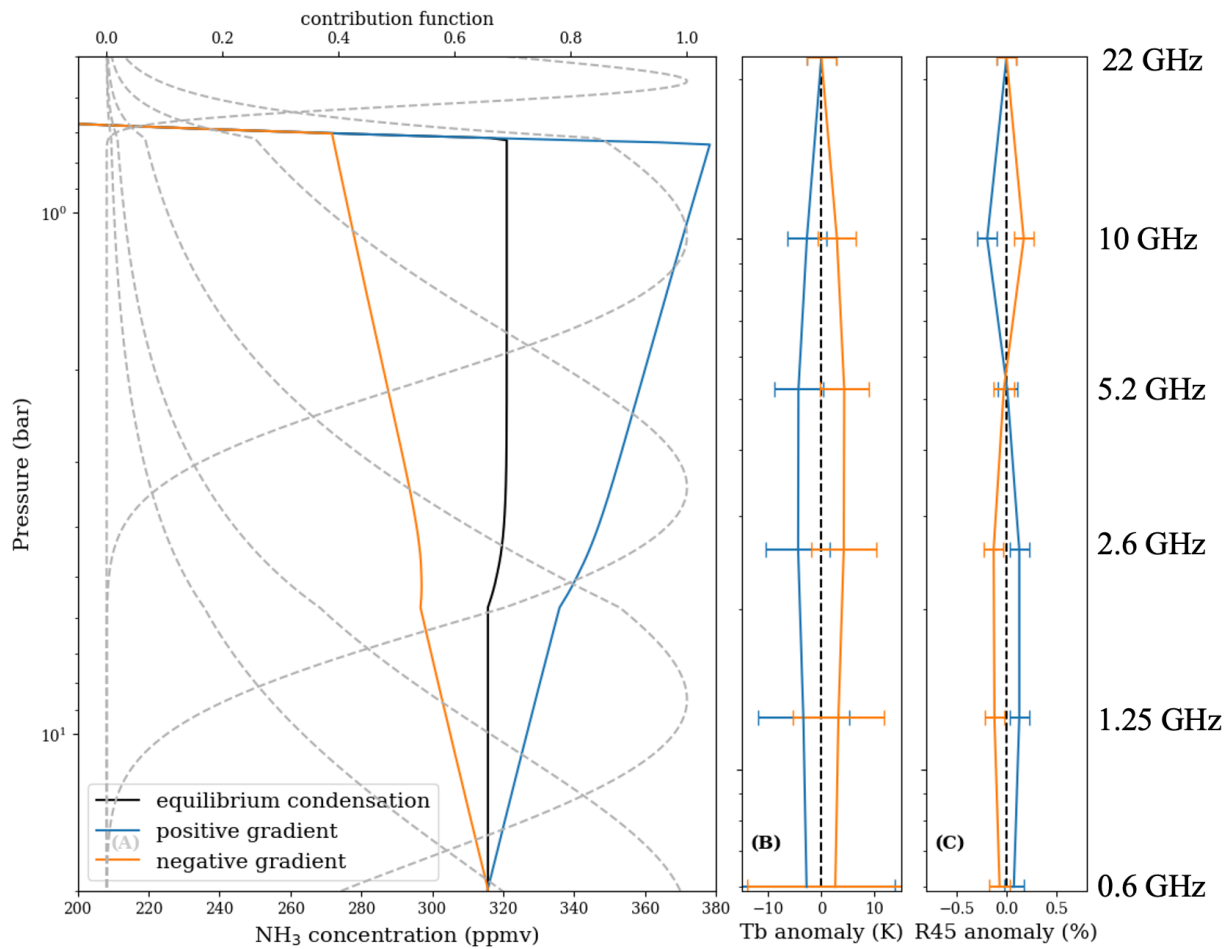

Figure 2. Similar to Figure 1 but for three Jovian ammonia profile models, equilibrium condensation (black), positive ammonia gradient with height (blue) and negative ammonia gradient (orange). The pressure level has been extended to 0.5 bar to show the gradient of ammonia vapor due to condensation and the contribution function of the 22 GHz channel.

The real atmosphere of Jupiter is far more complicated than the simple idealized model presented in Figures 1 and 2, and the inverse problem of atmospheric sounding is often termed "ill-posed" in inferring the atmospheric state using the given radiometric observations (Rodgers, 2000). This means that the solution is non-unique (there can be multiple possible solutions that could lead to the same spectra) and is subject to instability (small changes in the observed spectra could potentially lead to vastly different outcomes of the inferred atmospheric state). To address these issues, prior constraints must be incorporated into the inversion problem. They are essential additional pieces of information and assumptions that we use to bound the solution space and guide the solution toward a stable and objective answer with uncertainty quantification. The result is interpreted as the balance between the observational constraint and the prior constraint. Incorrect or overly restrictive prior constraints could potentially bias the solution or eliminate the true state from the feasible solution set.

In this article, we deem that the "adiabatic assumption" employed in previous radio inversions is too restrictive and that may prevent us from inferring the true condition of Jupiter's atmosphere, and by extension, those of other giant planets. As stated previously, the Galileo Probe detected a non-adiabatic temperature gradient, although in an anomalous "hot spot". The thermodynamics of Jupiter's weather layer present a complex system, mainly due to the phase transitions of water and ammonia. This phase transition results in the release of latent heat, which reduces the temperature gradient during a reversible adiabatic ascent (the moist adiabatic temperature gradient) (Holton, 1973). More importantly, a humid air parcel on Jupiter is heavier than a dry air parcel because water molecules are heavier than the background atmosphere of molecular hydrogen and helium. The gradient of water vapor caused by condensation and the subsequent change in the mean molecular weight result in the temperature gradient being potentially steeper than any dry adiabatic temperature profile across the water condensation layer (Friedson & Gonzales, 2017; T. Guillot, Chabrier, et al., 1994; T. Guillot, Gautier, et al., 1994; Leconte et al., 2017; Li & Ingersoll, 2015). This is a unique type of temperature profile in the Jovian atmosphere that does not exist on Earth. Despite this, the Jovian atmosphere can remain statically stable because the lower atmosphere (below the water cloud base) has a higher mean molecular weight than the upper atmosphere (above the water cloud) which has a lower mean molecular weight.

Thus, neither theoretical nor observational constraints compellingly support the assumption of strict adiabaticity for Jupiter's atmosphere. We therefore relax the "adiabatic" assumption in this work and proceed to utilize the possible super-adiabaticity in Jupiter's EZ to better constrain the water abundance in the deep atmosphere. This is possible because a super-adiabatic temperature gradient must be stabilized by a change of mean molecular weight, and water is the only condensable vapor that is abundant and heavy enough to change the mean molecular weight of the atmosphere significantly. Ammonia condensation could also alter the mean molecular weight of the atmosphere but by a much smaller amount, owning to its much lower abundance.

In what follows, we discuss the data processing steps in Section 2 and spectral inversion method and the prior constraints we used in the spectral inversion in Section 3, followed by the derived temperature and ammonia profiles with statistics in Section 4. In Section 5, we infer the deep-water abundance based on the inferred super-adiabatic temperature gradient. Section 6 concludes and summarizes the findings of this research. Finally, Section 7 discusses the caveats of this study and how it may be improved in the future. We also offer new insights on the physical implications of the Galileo Probe observation, the only in-situ measurement available for Jupiter's tropospheric temperature from 1 to 20 bars of pressure.

## 2 Juno MWR data processing

Jupiter's thermal emissions at six radio wavelengths were recorded from pole to pole at various emission angles by the Juno spacecraft while it scans the planet during each perijove encounter. The observed quantity at each microwave channel is the antenna temperature, which is the brightness temperature emitted by the atmosphere (or occasionally non-atmospheric source) convolved with the beam pattern of each antenna. The Juno MWR data processing pipeline removes the possible synchrotron contamination and determines a parameterized three-coefficient formula of the emission-angle-dependent brightness temperature of Jupiter's atmosphere at approximately 0.5-degree resolution in latitude (Oyafuso et al., 2020a). The formula is:

$$T_b(\mu) = c_0 - c_1 \frac{1-\mu}{1-\mu^2} + \frac{c_2}{2}\frac{(\mu - \mu^*)(1-\mu)}{(1-\mu^*)}, \tag{1}$$

in which $\mu$ is the cosine of the emission angle and, $\mu^* = 0.8$, corresponding to an emission angle of 37°. In this form, the coefficient $c_0$ measures the nadir brightness temperature; $c_1$ measures the absolute limb darkening (in unit K) at an emission angle of 37°; $c_2$ represents an additional reduction of the brightness temperature at $\mu = 0.6$ (53° emission angle) over a linear extrapolation by $c_0$ and $c_1$. Equation (1) has been validated over a variety of atmospheric conditions to adequately describe the emission angle dependence of the limb darkening curve from nadir looking up to 45° emission angle, while we find that the traditional two-parameter power-law approximation ( $T_b(\mu) = T_{b0}\mu^p$ ) is inadequate to describe the limb darkening at the precision of Juno MWR's observations (Oyafuso et al., 2020a).

Even for a homogenous atmosphere, the coefficients $c_0$, $c_1$ and $c_2$ change with latitude because the planet's effective gravitational acceleration is lesser at the equator and greater at the poles. For a given pressure interval, a larger gravitational acceleration yields a smaller photon path length and thus larger brightness temperature. The difference in the nadir brightness temperature ($c_0$) can be more than 10% between the equator and the pole in the 0.6-GHz channel. Before performing a spectral inversion, we (1) preprocess the Juno/MWR brightness temperatures by flagging and masking anomalies that may contain non-thermal emissions (such as auroral emissions) or large discrete features (such as the Great Red Spot) that appear only in one perijove and not the others; (2) project the observations from different latitudes to the equator by correcting for the effects of latitudinally varying gravitational acceleration; (3) obtain a globally averaged brightness temperature and limb darkening by averaging over the observations from Perijove 1 to Perijove 12; (4) contrast the equatorial observation (0 ~ 5 °N) with the global mean and obtain the equatorial anomaly with respect to the global mean. We focus on the Juno/MWR observations of Jupiter's atmosphere from Perijoves 1 to 12 because the first few orbits are the best to characterize Jupiter's equatorial region as the perijove latitude is closest to the equator. The resulting global-mean brightness temperature, limb darkening at a 45-degree emission angle and the equatorial anomaly with respect to the global mean are summarized in Table 1.

*Table 1. Globally averaged brightness temperatures, limb darkening and equatorial anomalies.*

| Frequency (GHz) | Global $T_b$ (K) | Global $R_{45}$ (%) | Equatorial $\delta T_b$ (K) | Equatorial $\delta R_{45}$ (%) |
|---|---|---|---|---|
| 0.6 | 841.1 ± 1.38 | 14.11 ± 0.11 | −39.4 ± 2.4 | 1.1 ± 0.1 |
| 1.25 | 451.6 ± 0.94 | 9.63 ± 0.056 | −33.4 ± 2.0 | −0.08 ± 0.1 |
| 2.6 | 325.9 ± 0.38 | 6.78 ± 0.037 | −28.3 ± 2.7 | 0.28 ± 0.16 |
| 5.2 | 244.6 ± 0.32 | 5.76 ± 0.024 | −22.4 ± 3.0 | −0.13 ± 0.16 |
| 10 | 189.6 ± 0.45 | 4.21 ± 0.041 | −12.7 ± 2.8 | −0.78 ± 0.25 |
| 22 | 140.6 ± 0.21 | 1.27 ± 0.036 | −3.1 ± 1.0 | 0.01 ± 0.29 |

Inspecting the last two columns of Table 1, one finds that (1) The EZ appears colder than the global mean at all microwave frequencies and (2) The 2.5-GHz channel shows a positive limb darkening anomaly (+0.28 %) and the 10-GHz channel shows a negative limb darkening anomaly (−0.78 %). These contrasts imply that the EZ is either enriched in ammonia vapor or its temperature structure is significantly different from elsewhere in Jupiter. In particular, the changing sign between the limb darkening values at 2.6-GHz and 10-GHz aligns with what is expected from a super-adiabatic temperature structure (Figure 1, blue line). However, the non-homogeneous distribution of ammonia can confound the inference of atmospheric temperature gradient due to the ammonia-temperature degeneracy. Thus, we would need to consider both the non-homogenous distribution of ammonia and the non-adiabatic temperature gradient for a comprehensive inversion of atmospheric state given Juno/MWR observations.

We aim to determine the potential temperature difference between the upper atmosphere and the deep interior of Jupiter, specifically investigating whether the temperature gradient follows a dry adiabat. To achieve this, we need two critical pieces of information.

The first is a value of the deep entropy of Jupiter's interior. This is conventionally expressed using potential temperature, denoted as $\theta_{1bar}$, which refers to the temperature that an air parcel *would reach* if dry-adiabatically decompressed to a 1-bar pressure level from some deeper level. Yet, given the significant variation of Jupiter's specific heat $c_p(T, f_{para})$ over the large range of kinetic temperatures between 1 and 300 bars and over the variation of the para-hydrogen fraction $f_{para}$ (Conrath & Gierasch, 1984), the conventional expression for potential temperature commonly adopted in meteorology (Emanuel, 1994) cannot be used here (Gierasch et al., 2004). Instead, for a diagnostic purpose, we present a definition of potential temperature in terms of small departure with respect to a dry adiabatic temperature profile, $T_{ad}(p)$. It follows that, for small departures, $\delta T(p)/T_{ad}(p) \approx \delta\theta(p)/\theta_0$, where $\theta_0$ is a reference constant, $\delta T(p)$ is the temperature anomaly and $\delta\theta(p)$ is the potential temperature anomaly. We demand that the deep atmosphere is isentropic, i.e. $\delta\theta(p) = 0$ everywhere. So, the deep isentrope is at $\theta = \theta_0$. To close approximation, $\theta(p) \approx \theta_0 + \delta\theta(p) = \theta_0 \frac{T(p)}{T_{ad}(p)} \equiv \tilde{\theta}(p)$, and we refer $\tilde{\theta}(p)$ the *adiabatically referenced potential temperature*. $\tilde{\theta}$ possess the following properties (1) a constant $\tilde{\theta}$ is a dry adiabatic temperature profile extended from the deep atmosphere. (2) $\tilde{\theta}$ directly reflects temperature profile. If $\tilde{\theta}$ decreases with height, the temperature profile is super-adiabatic. Conversely, if $\tilde{\theta}$ increases with height, the temperature profile is sub-adiabatic. (3) For an atmosphere with uniform heat capacity, $\tilde{\theta}$ converges to the conventional definition of $\theta$. We emphasize that $\tilde{\theta}$ only serves as a diagnostic purpose. When integrating the adiabatic temperature gradient, we consider the temperature dependence of the heat capacity and the variation of the para-hydrogen fraction. We have tested both equilibrium hydrogen and normal hydrogen and found negligible differences in the derived $\tilde{\theta}(p)$ because $T_{ad}(p)$ for these different choices is very nearly the same at pressures greater than ~ 3 bars. When assessing the stability of the atmosphere, we compare the density of two air parcels directly by lifting the air parcel at a higher pressure to the pressure level of the other air parcel. Due to the close relation between $\tilde{\theta}$ and $\theta$, from now on, we will omit the tilde symbol on top of $\tilde{\theta}$ and address the adiabatically referenced potential temperature as the potential temperature.

The second is the kinetic temperature at the 1-bar pressure level of Jupiter's atmosphere, denoted as $T_{1bar}$. If $T_{1bar} > \theta_{1bar}$, the atmosphere is sub-adiabatic; if $T_{1bar} < \theta_{1bar}$, the atmosphere is super-adiabatic, and if $T_{1bar} = \theta_{1bar}$, the atmosphere is dry adiabatic. Both $\theta_{1bar}$ and $T_{1bar}$ can be functions of latitude, meaning that different locations on Jupiter may have different static stability, as suggested by the general circulation model of giant planets (Schneider & Liu, 2009). From ground-based infrared observation, it is well established that $T_{1bar}$ is not uniform on Jupiter: the EZ is the coldest place and the North Equatorial Belt (NEB) is the warmest place (Fletcher et al., 2016; Gierasch et al., 1986). As the case with Earth's atmosphere, $T_{1bar}$ is modulated by atmospheric circulation and radiative processes in the upper troposphere and stratosphere (Li et al., 2018). Yet, $\theta_{1bar}$ (as used here to tag the reference for the deep isentrope) is generally regarded as uniform. This is because even a very small latitudinal difference in potential temperature along isobars would result in a vastly amplified thermal wind when integrated over the ~2000 km depth of the zonal circulation (Galanti et al., 2021). Although the longest wavelength channel is at 50 cm (0.6 GHz), which probes hundreds of bars of pressure, the 1.2-GHz channel (25 cm) is the best choice to determine the deep entropy because it is unaffected by the plasma absorption (Bhattacharya et al., 2023) and the ammonia concentration is less variable over latitude at the corresponding pressure range (30 ~ 50 bars) compared to shallower levels.

In summary, the 0.6-GHz channel is generally used for determining the abundance of alkali metals through the ionized electrons. The 1.2-GHz channel is for the deep ammonia concentration and the deep entropy. The 1.2 ~ 10-GHz channels are used to characterize the profiles of ammonia and temperature and the 22-GHz channel is used to determine the ammonia relative humidity near 0.5 bar pressure level.

## 3  Atmospheric profile inversion method

Almost all previous studies on the atmospheric profile inversion from radio observations assume a static, globally uniform and adiabatic temperature profile extrapolated downward from $T = 166\ K$ at the 1 bar pressure level (Li et al., 2017; Moeckel et al., 2023; de Pater et al., 2019). This is a fair assumption for low-precision measurements. As the precision of limb darkening reaches 0.1% and the precision of brightness temperature is less than 1 K, a globally uniform and adiabatic temperature structure is no longer appropriate. As demonstrated in Figure 1, a 0.15 K/km non-adiabatic temperature yields a detectable signal in both nadir brightness temperature and limb darkening observations. Thus, we allow both the kinetic temperature and ammonia concentration to vary in the vertical.

To leverage the stability of the Juno/MWR observation, meaning that the relative precision is much higher than the absolute accuracy, we employ a two-stage differential fitting method, as has been done in fitting Saturn's radio observation using the Very Large Array (VLA) (Li et al., 2023). The first stage fits the global mean spectra of Jupiter's atmosphere based on absolute values of brightness temperature and limb darkening (assuming 2% calibration uncertainty and 0.5 K measurement noise) and the second stage fits the equatorial anomaly based on the differential observation shown in the last two columns of Table 1. The theory of differential fitting is derived in Li et al. (2023) for the radio observation of Saturn's atmosphere and we use it here for the observation of Jupiter.

The fitting in the first stage establishes the value of the deep entropy, $\theta_{1bar}$ and the deep ammonia concentration, $X_{NH3}$ because these two quantities are assumed to be globally uniform. Before the Juno mission, the Galileo Probe in-situ measurement and the Voyager radio occultation experiments were used to estimate the value of $\theta_{1bar}$. Assuming a dry adiabatic thermal structure, the deep entropy is $\theta_{1bar} = T_{1bar} = 166.1 \pm 0.8\ K$ (Seiff et al., 1998; Young et al., 1996). The concentration of ammonia on Jupiter has also been directly measured by the Galileo Probe mass spectrometer ($566 \pm 216$ ppmv) (Wong et al., 2004), indirectly inferred by the radio attenuation of the Probe signal ($700 \pm 100$ ppmv) (Folkner et al., 1998) and also from the preliminary spectral inversion from the Juno/MWR ($362 \pm 33$ ppmv) based on nadir brightness temperature only (Li et al., 2017). Curiously, the nominal value obtained by the previous Juno/MWR inversion is inconsistent with the lower limit of the value obtained by the Galileo Probe mass spectrometer. Such a discrepancy can be attributed to an inverted ammonia gradient (higher ammonia concentration at higher altitudes) found by the ammonia profile retrieval near Jupiter's equator (Li et al., 2020; Moeckel et al., 2023). Consequently, previous Galileo Probe measurements cannot provide a reliable estimate of $\theta_{1bar}$ or $X_{NH3}$ if we allow a non-adiabatic temperature gradient and accept the inverted ammonia gradient near the equator (Section 7 discusses the intricacies of interpreting the Galileo Probe observation). We will regard these two parameters as unknown and estimate them from Juno/MWR observations summarized in Table 1.

The fitting in the second stage establishes the temperature gradient via differential fitting. Differential fitting bypasses the calibration uncertainty and measures atmospheric variability. Similar to (Li et al., 2023), The $\chi^2$ of the differential fitting for each MWR channel is defined as:

$$\chi^2 = \left(\frac{\delta T_b^{model} - \delta T_b^{obs}}{\sigma(\delta T_b^{obs})}\right)^2 + \left(\frac{\delta R_{45}^{model} - \delta R_{45}^{obs}}{\sigma(\delta R_{45}^{obs})}\right)^2, \tag{2}$$

where $\delta T_b^{model} = T_{b,eq}^{model} - T_{b,global}^{model}$ represents the modeled difference in nadir brightness temperature between the EZ and the global mean; $\delta R_{45}^{model} = R_{45,eq}^{model} - R_{45,global}^{model}$ represents the modeled difference in limb darkening at a 45° emission angle between the EZ and the global mean. $\delta T_b^{obs}$ and $\delta R_{45}^{obs}$ are the observed brightness temperature and limb darkening difference between the EZ and the global mean, where $T_b$ and $R_{45}$ come from radiative transfer calculation given the profiles of temperature and ammonia concentration. Differential fitting is useful for contrasting two locations, such as the global mean and the EZ. Later, we will gather the spectral inversion statistics and demonstrate that EZ has a steeper temperature gradient than the global mean.

There are several ways to perform spectral inversion and the choice is based on the type of measurement and the sensitivity of the instrument. For example, spectral inversions for radio observations of giant planets using VLA were often done by hand-picking several vertical levels and adjusting the ammonia concentration manually (de Pater et al., 2001, 2016, 2019; Sault et al., 2004). As the spatial resolution, spectral resolution and instrument precision improves, parameterized inversions were adopted (Li et al., 2020; Moeckel et al., 2023) in studying Jupiter's ammonia concentration using Juno/MWR observations. In particular, (Li et al., 2020) assumed that the ammonia profile $q(p)$ at Jupiter's equator behaves like the functional form of:

$$q(p) = A + B \exp\left(-\frac{p - p_D}{\Delta p}\right), \tag{3}$$

and invert for the four parameters, $A, B, p_D$ and $\Delta p$. Moeckel et al. (2023) parameterized the ammonia profile $q(p)$ as:

$$q(p) = q^{deep} - H_{mix}(\ln P_b - \ln P), \tag{4}$$

and invert for the three parameters, $q^{deep}, H_{mix}$ and $P_b$.

Equations (3) and (4) are restrictive in the representation of a realistic ammonia profiles, because any ammonia profile that does not follow the functional form of (3) or (4) will be excluded from the solution set and thus bias our solution of atmospheric structure. Yet, it is impossible to enumerate all functions because the atmospheric profile has an infinite degree of freedom. The most sophisticated model constructs a continuous atmospheric profile level-by-level as has been done in Li et al. (2017). Here, we further improve the way of constructing a smooth atmospheric profile by Gaussian Process, which yields smooth atmospheric profiles with better uncertainty quantification than a piecewise linear construction.

Let $Z$ be the log-pressure coordinate defined as:

$$Z = H_0 \log\left(\frac{p_0}{p}\right), \quad (5)$$

Where $H_0$ is a reference atmosphere scale height and $p_0$ is the reference pressure such that $Z = 0$ at $p = p_0$. A uniform grid in $Z$ means that the pressure is sampled uniformly in log space. Let $T(Z)$ be the vertical temperature anomaly and $\text{NH}_3(Z)$ be vertical ammonia anomaly. We sample $N$ layers of the atmosphere and connect them via a Gaussian interpolation process. Specifically, we use $N = 8$ in our spectral inversion with maximum and minimum pressure boundaries at 100 and 0.5 bars, respectively. The sampling pressure levels are at 100, 46.9, 22.0, 10.3, 4.84, 2.27, 1.07, and 0.5 bars respectively. These numbers are guided by the number of microwave channels, the spread of their weighting functions and the density scale height in the physical space. Choosing 8 levels yields roughly one sampling point per scale height. No improvement was found when greater than 10 levels is used.

Denoting the $N = 8$ levels at $Z^* = \{Z_i^* \mid 1 \leq i \leq N\}$, the Gaussian Process assumes that $T_i^* = T(Z_i^*)$ and $X_i^* = \text{NH}_3(Z_i^*)$ form a $N$-dimensional Gaussian distribution respectively. The covariance of the multidimensional Gaussian is described by a kernel function, which we choose to be the squared exponential kernel, defined as:

$$k_{SE}(Z_i, Z_j) = \sigma^2 \exp\left(-\frac{(Z_i - Z_j)^2}{2L^2}\right), \quad (6)$$

where $L$ is the correlation length between layers and $\sigma$ is the prior standard deviation. The correlation length $L$ is used to model the correlation between adjacent levels and suppresses structures finer than the correlation length $L$. Another popular kernel function is the exponential kernel, defined as:

$$k_E(Z_i, Z_j) = \sigma^2 \exp\left(-\frac{|Z_i - Z_j|}{L}\right). \quad (7)$$

This kernel function is the result of a random walk model with uncorrelated Gaussian random noise (Rodgers, 2000).

Given the N pairs of $\{(Z_i^*, T_i^*) \mid 1 \leq i \leq N\}$ and $L$, we construct a smooth profile of $T(Z)$ that passes through each $(Z_i^*, T_i^*)$. Letting $Z = \{Z_i \mid 1 \leq i \leq M\}$ be the vertical grid of the atmospheric model, we define the following two covariance matrices:

$$S(Z, Z^*)_{i,j} = k_{SE}(Z_i, Z_j^*)$$
$$S(Z^*, Z^*)_{i,j} = k_{SE}(Z_i^*, Z_j^*)$$
(8)

Let $\boldsymbol{T^*}$ be the vector of $\{T(Z_i^*) \mid 1 \leq i \leq M\}$ and $\boldsymbol{T}$ be the vector of $\{T(Z_i) \mid 1 \leq i \leq N\}$. Vector $\boldsymbol{T}$ at all $Z_i$ levels is calculated by the interpolation equation:

$$\boldsymbol{T} = S(Z, Z^*) S(Z^*, Z^*)^{-1} \boldsymbol{T^*},$$
(9)

which is a simplified version of a more general Gaussian Process assuming that an uninformed state is at zero anomaly. Finally, we set the lower and upper boundary conditions by stipulating that the values at pressure levels beyond the sampling range should be the same as the value at the nearest sampling level.

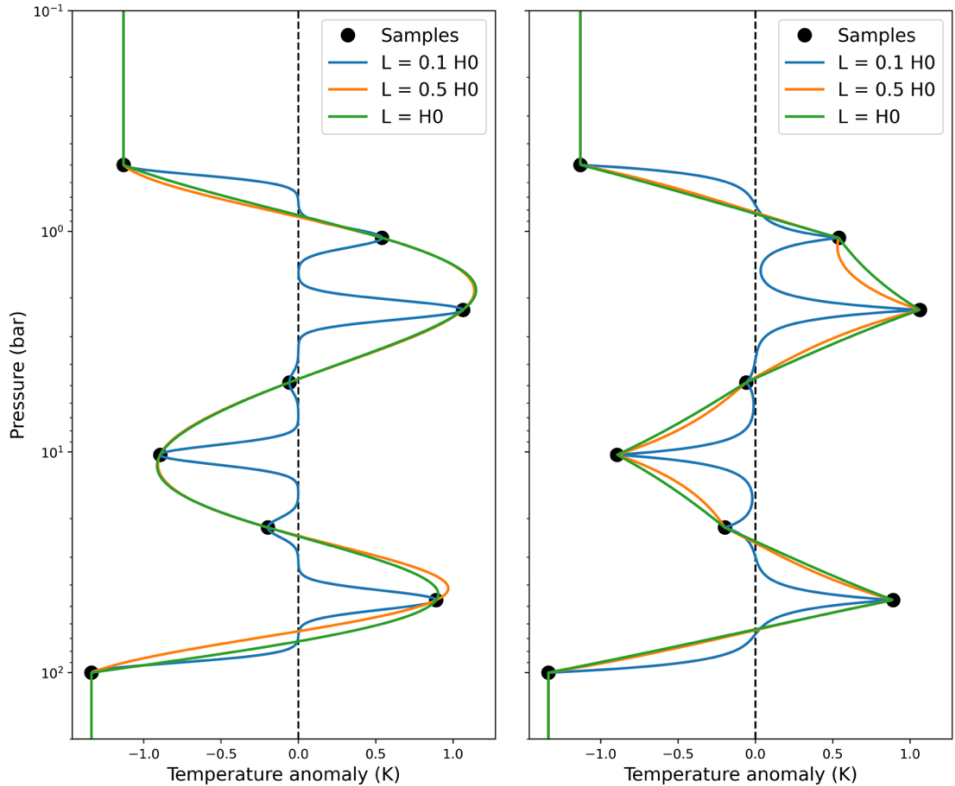

*Figure 3. Smooth atmospheric profile constructed from discrete samples. The left panel shows the profile constructed by the squared exponential kernel and the right panel shows the exponential kernel. Black dots show the temperature anomalies sampled at 8 levels. The blue, orange and green lines are constructed atmospheric profiles given three different values of correlation length $L = 0.1\ H_0, 0.5\ H_0$ and $H_0$ respectively. The separation between sampling levels is roughly $H_0$.*

Figure 3 illustrates the effect of correlation length and the choice of kernel function. The left panel shows the result of the squared exponential kernel and the right panel is the result of using the exponential kernel. In both cases, when the correlation length is small ($L = 0.1\ H_0$), in between two sampled pressure levels, the interpolated anomaly profile drops toward zero. At longer correlation lengths ($L = 0.5\ H_0, H_0$), the profile is smoothly interpolated in between. The exponential kernel converges to a piecewise linear interpolation, with interpolated values to be always in between the sampled values, thereby disallowing any interior maximum. In contrast, the squared exponential kernel allows for a much smoother interpolation, where the interpolated values can form a smooth curve that may exceed the sampled values. Since it is unknown where the maximum temperature or ammonia concentration may occur in the atmosphere, the squared exponential kernel yields profiles that are less sensitive to the selection of sampling levels while the exponential kernel can lead to significant bias when the sampling levels does not correspond to the location of maximum temperature or ammonia anomaly.

The above interpolation procedure utilizes random process (RP) to map a set of discrete values to a continuous function, denoted as:

$$I_{RP}: \boldsymbol{T}^* \to \boldsymbol{T} \tag{10}$$

Strictly speaking, RP is statistical, meaning that a range of profiles are generated with a mean and a variance. We simplified the procedure by only using the mean and ignoring the variance. By doing so, the likelihood estimation given the mean of the constructed continuous function is the same as the likelihood estimation given the discrete samples, i.e.

$$P(\boldsymbol{Y} \mid \boldsymbol{T}) = P(\boldsymbol{Y} \mid \boldsymbol{T}^*), \tag{11}$$

where vector $\boldsymbol{Y}$ is the observed brightness temperatures at multiple frequencies and emission angles. $P(\boldsymbol{Y} \mid \boldsymbol{T})$ is a statistical notation denoting the probability density of observing $\boldsymbol{Y}$ given $\boldsymbol{T}$.

Equation (11) is valid when a set of discrete samples $T^*$ uniquely maps to a continuous function $T$. So, in the following sections, we will ignore the statistical essence of $I_{RP}$ and regard $I_{RP}$ as a deterministic interpolation function that yields the statistical mean of the RP, i.e.:

$$I_{RP}: T^* \rightarrow T = E(T \mid T^*). \tag{12}$$

A similar process applies to constructing the ammonia profile $X$, in which the interpolation give a continuous ammonia profile $X$ given a discrete set of samples $X^*$.

In the case of interpreting Juno/MWR observations, let $Y$ be the observed brightness temperatures/limb darkening at multiple frequencies. Let $S_Y$ be the covariance matrix of observation $Y$ and $F$ be the forward radiative transfer model that takes in the atmospheric profile $(T, X)$ and outputs the angular-dependent brightness temperatures. We define the $\chi^2$ parameter of the atmospheric profiles $T, X$ as:

$$\chi^2 = (Y - F(T, X))^T S_Y^{-1} (Y - F(T, X)), \tag{13}$$

which is a general form of equation (2) that measures the goodness of fit to the observation. Assuming Gaussian statistics, the probability distribution function of observing $Y$ given the profiles $T, X$ is:

$$P(Y \mid T, X) = P(Y \mid T^*, X^*) = (2\pi)^{-M/2} |S_Y|^{-1/2} \exp(-\chi^2/2), \tag{14}$$

where $M$ is size of $Y$. According to Bayes' theorem, the joint probability distribution function of the discrete samples $T^*, X^*$ given the observation $Y$ is thus:

$$P(T^*, X^* \mid Y) = \frac{P(Y \mid T^*, X^*) P(T^*, X^*)}{P(Y)}, \tag{15}$$

which means that the probability of an atmosphere to have sampled values $T^*, X^*$ given the observation $Y$ equals the probability of having $Y$ observed given sampled values $T^*, X^*$ multiplied by the prior probability of $T^*, X^*$ and normalized by the probability of observing $Y$. Since we sample temperature and ammonia profiles independently, the joint probability is:

$$P(T^*, X^*) = P(T^*) P(X^*). \tag{16}$$

The term $P(T^*)$ in equation (16) is given by the Gaussian statistics:

$$P(T^*) = (2\pi)^{-N/2} |S(z^*, z^*)|^{-1/2} \exp\left(-\frac{1}{2}(T^{*T} S(z^*, z^*)^{-1} T^*)\right), \tag{17}$$

and $P(X^*)$ is obtained in a similar manner.

We use the Markov Chain Monte Carlo (MCMC) method to perform the statistical inference of $T, X$ given Juno/MWR observation $Y$. We employ the Affine Invariant Ensemble Sampler algorithm to conduct the sampling (Goodman & Weare, 2010). We use 64 parallel MCMC walkers to explore the parameter space. Each walker takes 2000 steps, and the final statistics are collected from the positions of the last 1000 steps taken by all walkers.

The MCMC chain is initialized by randomly drawing discrete temperature $T^*$ and ammonia anomalies $X^*$ at the sampling pressure levels. The top sampling level is at 0.5 bar and the bottom level is at 100 bar. Below the 100-bar level, we assume the atmosphere to be isentropic with constant ammonia and water mixing ratio. Above the 0.5 bar level, where the Juno MWR has no sensitivity, the atmosphere is assumed to be isothermal at the temperature of the upper boundary. The prior probability of $P(T^*, X^*)$ is evaluated by equations (16) and (17). The choice of the prior standard deviations of temperature and ammonia depends on whether we perform an absolute fit or a differential fit. To fit the global mean spectra (absolute fit), we use 5 K as the prior standard deviation of temperature and 50 ppmv as the standard deviation for ammonia concentration. The prior standard deviation for the temperature is informed by the latitudinal temperature variability observed in mid-infrared measurements (e.g. Fletcher et al., 2016; Gierasch et al., 1986). The prior standard deviation for ammonia concentration is estimated by experimentation, i.e. increasing the prior standard deviation until a satisfactory fit to the MWR data is achieved. In fitting the equatorial anomaly (differential fit), we use 20 K as the prior standard deviation of temperature and 500 ppmv as the standard deviation for ammonia concentration. These values are significantly larger than those used in the first stage because the equatorial anomaly exhibits a more pronounced difference compared to the global mean.

Then we construct continuous atmospheric profiles $T$ and $X$ based on discrete $T^*, X^*$ using $I_{RP}$. After constructing the atmospheric profiles, it is crucial to ensure that the constructed temperature and ammonia profiles are physically possible. Some MCMC models assign zero prior probability to unphysical states. However the method does not apply to inferring the state of the Jovian troposphere because there are orders of magnitude more unphysical states than physical states.

Supposing that the Jovian troposphere is nearly adiabatic and we wish to infer the small deviation from adiabaticity, the stability criteria demand a monotonic increasing potential temperature with height (neglecting the contribution of water in this example). If we randomly sample the potential temperatures at $N = 8$ layers in the atmosphere, the chance to have monotonically increasing values is $2^{-7} \approx 0.78\%$. So, assigning zero prior probability to unphysical states would reject more than 99% samples. Therefore, we opt to rectify the constructed profiles based on constraints imposed by stability and thermodynamic considerations after $I_{RP}$. The step is denoted by:

$$R_{TX}: (\boldsymbol{T}, \boldsymbol{X}) \to (\widetilde{\boldsymbol{T}}, \widetilde{\boldsymbol{X}}), \tag{18}$$

where $\widetilde{\boldsymbol{T}}, \widetilde{\boldsymbol{X}}$ are rectified physical states from proposed states $\boldsymbol{T}, \boldsymbol{X}$ (may be unphysical).

$R_{TX}$ entails the following three sub-steps: (1) examining the thermal and compositional profiles layer-by-layer, starting from the bottom and moving upwards, (2) correcting for any statically unstable layer to neutral stability, and (3) removing excess vapors that exceed the saturation vapor pressure. $R_{TX}$ resembles the convective adjustment process developed for GCMs (Arakawa & Schubert, 1974; Betts & Miller, 1993) or single-column models (Manabe & Wetherald, 1975). However, unlike these models, we directly alter the concentration of condensable vapors and the temperature to satisfy the stability criteria, without considering time scales, mass conservation, or enthalpy conservation. To account for the changing of the mean molecular weight of the atmosphere across the water condensation level, we prescribe a profile of water vapor according to its saturated profile. It is important to note that while water is included in the analysis, it is not part of the inversion solution. Instead, it merely controls the degree of super-adiabaticity of the temperature profiles. A zero-water abundance would prevent the occurrence of any super-adiabatic temperature gradient. Conversely, detecting the presence of any super-adiabatic temperature gradient implies the presence of water, and thus the ability to infer water abundance. This concept is further elaborated in Sections 4 and 5. Figure 4 illustrates temperature profiles constructed by $I_{RP}$ and the subsequent correction step for three randomly sampled temperature anomaly profiles. The local peak signifies the sampling pressures. The dashed line represents the profile derived from $I_{RP}$, while the solid lines illustrate the rectified profiles. Neutrally stratified atmospheric layers are indicated by vertical segments in the virtual potential temperature (Emanuel, 1994).

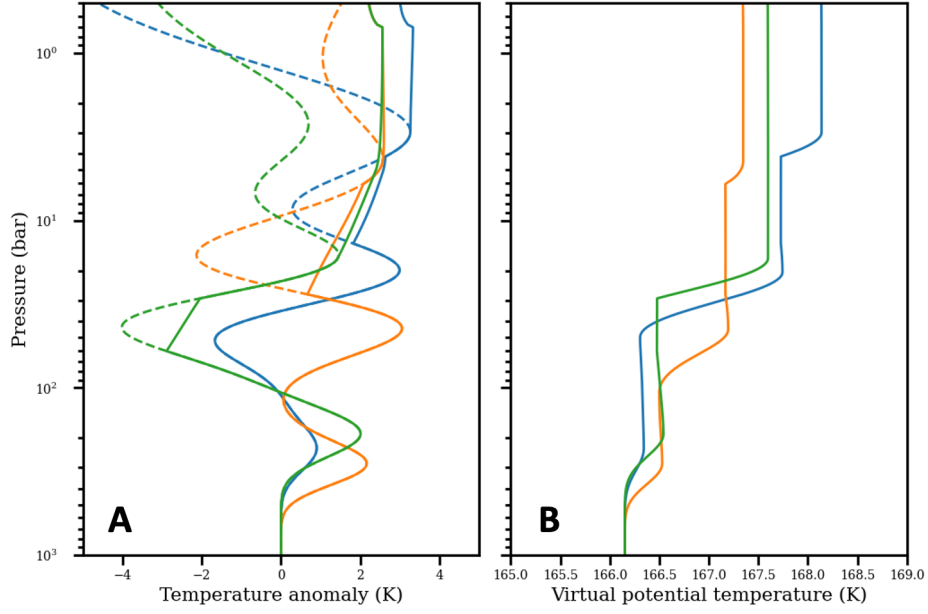

*Figure 4. Three randomly sampled temperature profiles (green, blue, orange). (A) Dashed lines show the non-adiabatic temperature anomaly constructed according to equation (9). Solid lines show adjusted temperature profiles with unstable layers removed. (B) Virtual potential temperature profiles of the three samples. A vertical line segment of the virtual potential temperature indicates neutral stability.*

Note that, altering $T, X$ after $I_{RP}$ but before calculating the likelihood function is allowed in the MCMC sampling. This is because the mapping from $(T, X) \to (\widetilde{T}, \widetilde{X})$ is unique. Therefore, the following construction chain

$$(T^*, X^*) \xrightarrow{I_{RP}} (T, X) \xrightarrow{R_{TX}} (\widetilde{T}, \widetilde{X}) \tag{19}$$

yields a unique mapping from $(T^*, X^*)$ to $(\widetilde{T}, \widetilde{X})$ and thus:

$$P(Y \mid \widetilde{T}, \widetilde{X}) = P(Y \mid T, X) = P(Y \mid T^*, X^*). \tag{20}$$

After getting $\widetilde{T}, \widetilde{X}$, a forward radiative transfer calculation $F(\widetilde{T}, \widetilde{X})$ is carried out to yield the likelihood $P(Y \mid \widetilde{T}, \widetilde{X})$ of matching observation $Y$. The radiative transfer model is the JAMRT model that calculates the opacity from $H_2O$, $NH_3$, $H_2S$, $H_2$-$H_2$/$H_2$-He collisional induced absorption and free-free electron absorption (Bellotti et al., 2016; Hanley et al., 2009; Janssen et al., 2017; Li, Le, et al., 2018; Oyafuso et al., 2020b). But we did not include water cloud given that the opacities are low, based on the derived cloud bulk densities and the measured microwave properties of aqueous ammonia (Duong et al., 2014). Finally, the posterior probability $P(\widetilde{T}, \widetilde{X} \mid Y)$ is derived using the Bayes' law. Combing the above steps, we have the following inference sequence:

$$(T^*, X^*) \xrightarrow{I_{RP}} (T, X) \xrightarrow{R_{TX}} (\widetilde{T}, \widetilde{X}) \xrightarrow{F, Y} P(Y \mid \widetilde{T}, \widetilde{X}) \xrightarrow{\text{Bayes}} P(\widetilde{T}, \widetilde{X} \mid Y). \quad (21)$$

The above sequence suggests that each state $(T^*, X^*)$ yields a unique posterior probability $P(\widetilde{T}, \widetilde{X} \mid Y)$. Since we cannot directly infer $P(T^*, X^* \mid Y)$ without using a continuous atmospheric profile, we treat $P(\widetilde{T}, \widetilde{X} \mid Y)$ as an approximate for $P(T^*, X^* \mid Y)$ as has been done in various retrieval studies for the Jovian stratosphere in the infrared (e.g. Fletcher et al., 2010, 2016). Let $(T^*, X^*)^n$ be the state at the $n$-th step, the MCMC algorithm proposes a new state $(T^*, X^*)^{n+1}$ for the $(n + 1)$-th step. Either $(T^*, X^*)^{n+1}$ is added to the chain or $(T^*, X^*)^n$ is added depending on their relative posterior probabilities. The chain is then advanced from $n$ to $n + 1$, and the steps are repeated until states in the chain reach equilibrium. After the chain converges, we infer the posterior distribution given all states in the chain. Each state $(T^*, X^*)$ in the chain is associated with a vertical profile pair $(\widetilde{T}, \widetilde{X})$ and a posterior probability $P(T^*, X^* \mid Y)$. The final mean profile is the weighted average of all profiles in the chain with the weight being $P(T^*, X^* \mid Y)$. The standard deviation of the inferred profile is obtained in a similar manner.

## 4 The derived temperature and ammonia vapor profile

Figure 5 presents four different Juno MWR retrieved kinetic temperature profiles from 1 bar to 20 bars near the equator (0 ~ 5 °N). As discussed in Section 3, the deep water abundance is not part of the inversion variable. We manually experiment with four different deep-water abundances, 0.1, 1, 2, and 4 times solar (solar oxygen abundance according to Asplund et al., 2009) to elucidate the effect of water on the temperature profile and the goodness-of-fit to Juno/MWR data. The Galileo Probe profile, representing an adiabatic fit to the ASI data, matched to 260K at 4.18 bar (Seiff et al., 1998) is drawn in parallel for comparison.

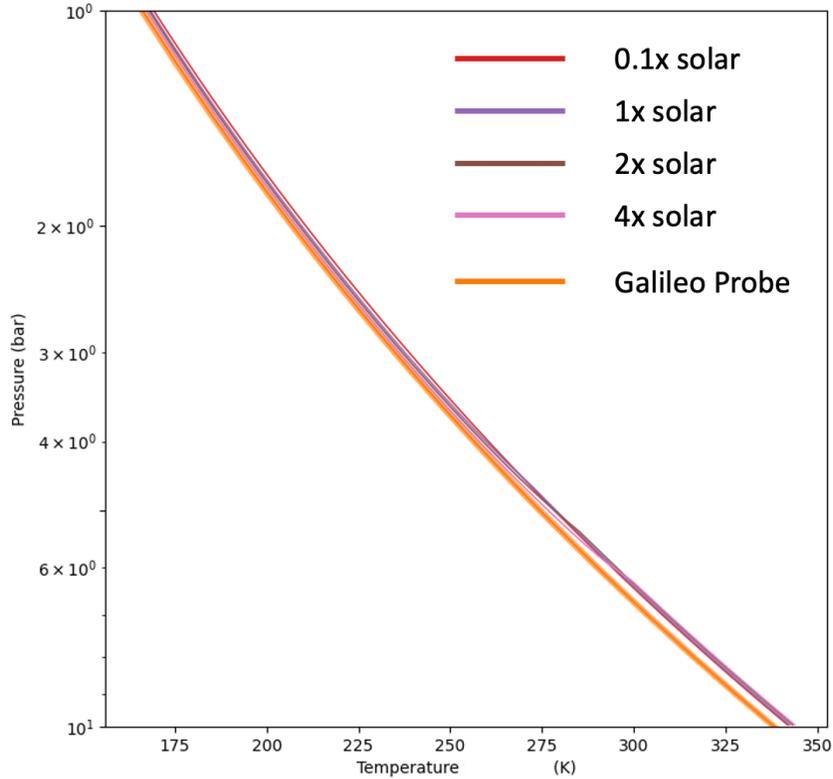

*Figure 5. Inferred temperature-difference profiles for four deep-water abundances. The uncertainties of the temperature profiles are too small to discern at this scale.*

We observe that, regardless of the choice of the deep water abundance, the Juno/MWR-derived temperature profile is unambiguously warmer than the Galileo Probe temperature profile at pressures greater than 5 bars. This means that Jupiter's deep entropy $\theta_{1bar}$ is larger than the kinetic temperature at the 1-bar pressure level, $T_{1bar}$. Figure 6 zooms in the difference between the temperature profiles to show the uncertainty of the profile inversion. The uncertainty of the Galileo Probe temperature profile is about $\pm 1$ K and the uncertainty of the Juno MWR derived temperature is generally at about $\pm 2$ K at all pressure levels.

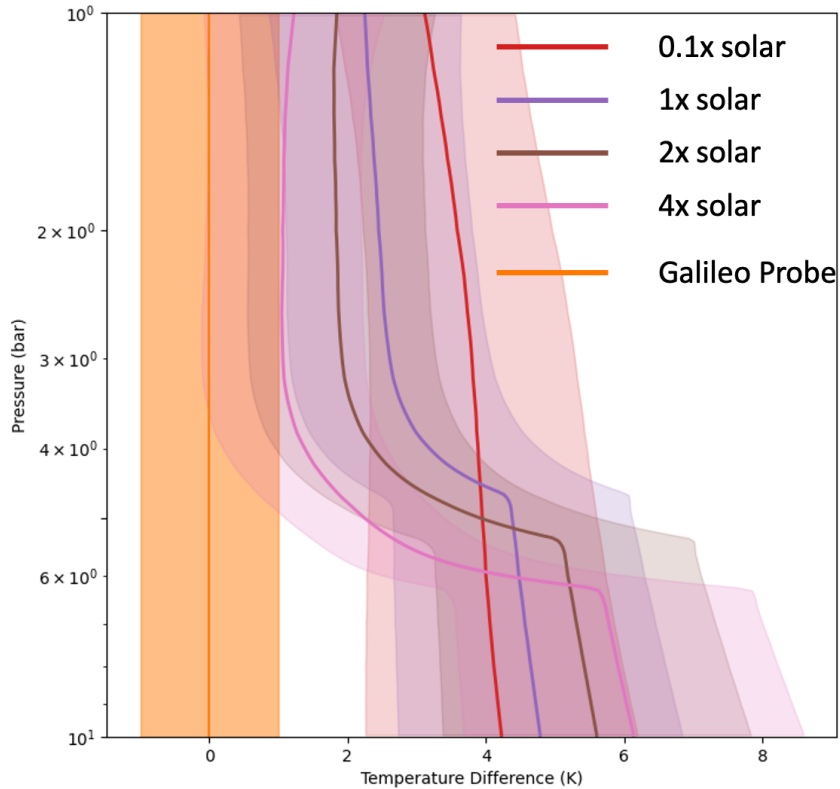

*Figure 6. Difference between temperature profiles and uncertainties. The shaded area depicts the uncertainty of each case, which is labeled on the top right corner. The Galileo Probe profile is a reference, so it appears as a vertical line in the figure.*

Comparing the global mean profile with the equatorial profile at various deep water abundances reveals the impact of water on the temperature profile. Figure 7A compares the case with zero water abundance at the equator to the global mean. The equatorial deep entropy diverges from the global-mean deep entropy. The agreement improves as the water abundance increases, as shown in Figure 7B-D, with the equatorial deep entropy matching the global isentrope the best when the deep water abundance is four times solar. This is because the EZ appears "cold" in brightness temperature and one possible scenario is that the EZ is indeed colder between some pressure levels. If there is no water in the EZ, the only way to reduce the temperature is to reduce the temperature of the entire atmosphere since the shallow atmosphere is connected to the deep atmosphere via an adiabatic relation.

However, when the water abundance is greater than zero, a super-adiabatic temperature gradient is permitted across the water condensation level, and the super-adiabaticity increases with the deep-water abundance. The development of a super-adiabatic temperature allows a new temperature solution that is cold at the shallow atmosphere (as a fitting to the "cold" brightness temperature) and warm at the deep atmosphere (as a fitting to the limb darkening anomaly). Collectively, they bring the equatorial deep entropy close to the global mean deep entropy. Further increasing the deep-water abundance does not improve the fitting. In all cases with a finite value of the deep-water abundance, $T_{1bar}$ is lower than $\theta_{1bar}$. In the case of four times solar water abundance, $\theta_{1bar}$ is approximately $169 \pm 1.6$ K while $T_{1bar}$ is $166 \pm 1$ K as determined by the Galileo Probe.

In addition to the non-adiabatic gradient of temperature, a gradient of ammonia vapor is also found at the equator, as has already been discussed (Li et al., 2020; Moeckel et al., 2023). In all cases of the deep water abundance, the ammonia vapor exhibits a positive gradient (cyan lines in Figure 7). The peak ammonia concentration reaches 500 ppm in between 1 and ~ 3 bars, a level consistent with the Galileo Probe measurements. Then, the ammonia concentration decreases with increasing depth, reaching $315 \pm 20$ ppm at depth. The deep ammonia concentration determined by this study is less than that reported in (Li et al., 2017) because we have improved our radiative transfer model by considering the opacity contribution from thermally ionized electrons from alkali metals (Bhattacharya et al., 2023).

One caveat of this inversion method is that the super-adiabatic temperature gradient only develops at the water condensation level near a few bars, because the profile of water is prescribed to follow the saturation curve. In reality, a gradient of water vapor may exist anywhere in the deep atmosphere, as imposed for example by the large scale circulation, and the water vapor may be sub-saturated, but the inversion method does not permit consideration of these possibilities. Yet, a robust result emerging from these trial values of the deep water abundance is the existence of a super-adiabatic temperature gradient.

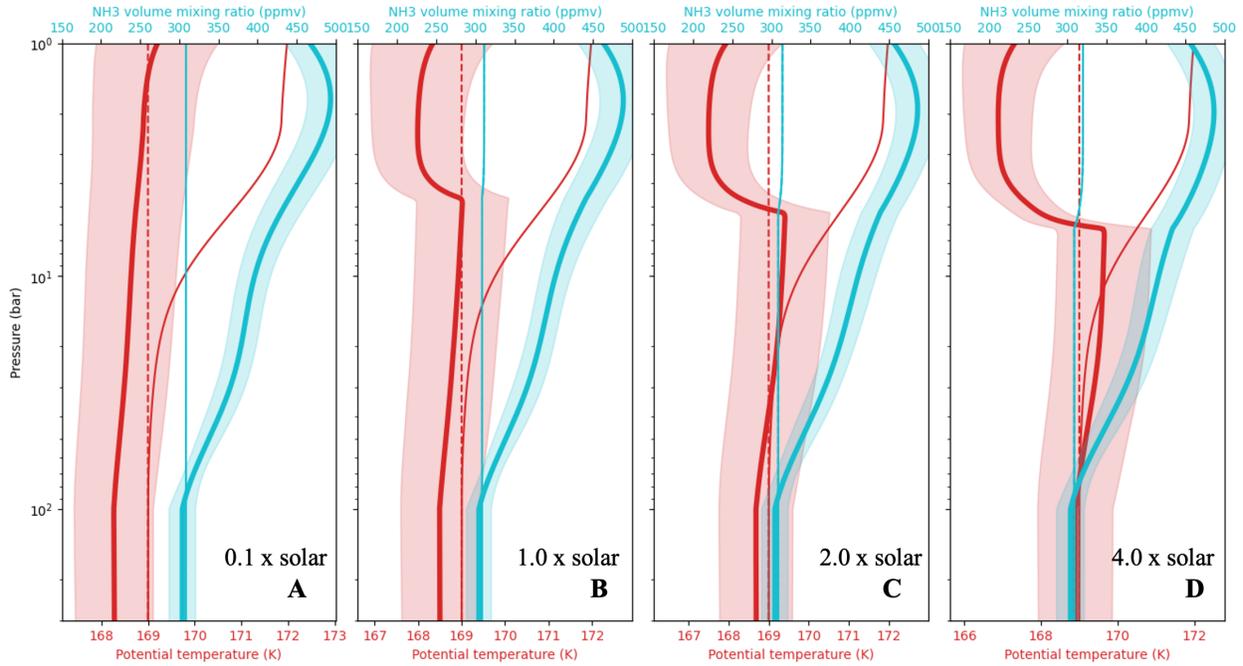

*Figure 7. Global and equatorial profiles. Thin red lines: global mean temperature profile. Thick red lines: equatorial temperature profile. Red shades: temperature profile inversion uncertainty. Thin cyan lines: global ammonia profile. Thick cyan lines: equatorial ammonia profile. Cyan shades: ammonia profile inversion uncertainty. Dashed lines: reference dry adiabatic profile. The deep water abundance increases from left to right, as indicated in the figur .*

Figure 8 compares the fitting of two cases: a dry case (blue contours) and a four-times-solar-water case (orange contours). By inspecting panels E and F in Figure 8, it is evident that the low brightness temperature observed in the 10 GHz and 22 GHz channels results in the super-adiabatic temperature gradient. This is consistent with what we have found in the synthetic study in Figure 1. Since the inversion model freely evolves its temperature at the 1-bar pressure level, there is a degeneracy between increasing the ammonia abundance and decreasing the temperature at a few bars of pressure. The ammonia concentration in the dry case is higher than in the four-times-solar-water case to compensate for the low brightness temperature. However, the four-times-solar-water case fits the Juno/MWR data at 0.6, 5, 10, and 22 GHz channels better because the clustering of points is closer to the Juno/MWR observation (black crosses). The improved fit at the 0.6-GHz channel for the four-times-solar-water case is due to the increased water opacity at depth, which becomes greater than ammonia opacity and that contributes to a better fit. The improved fits in other channels are driven by a super-adiabatic temperature profile.

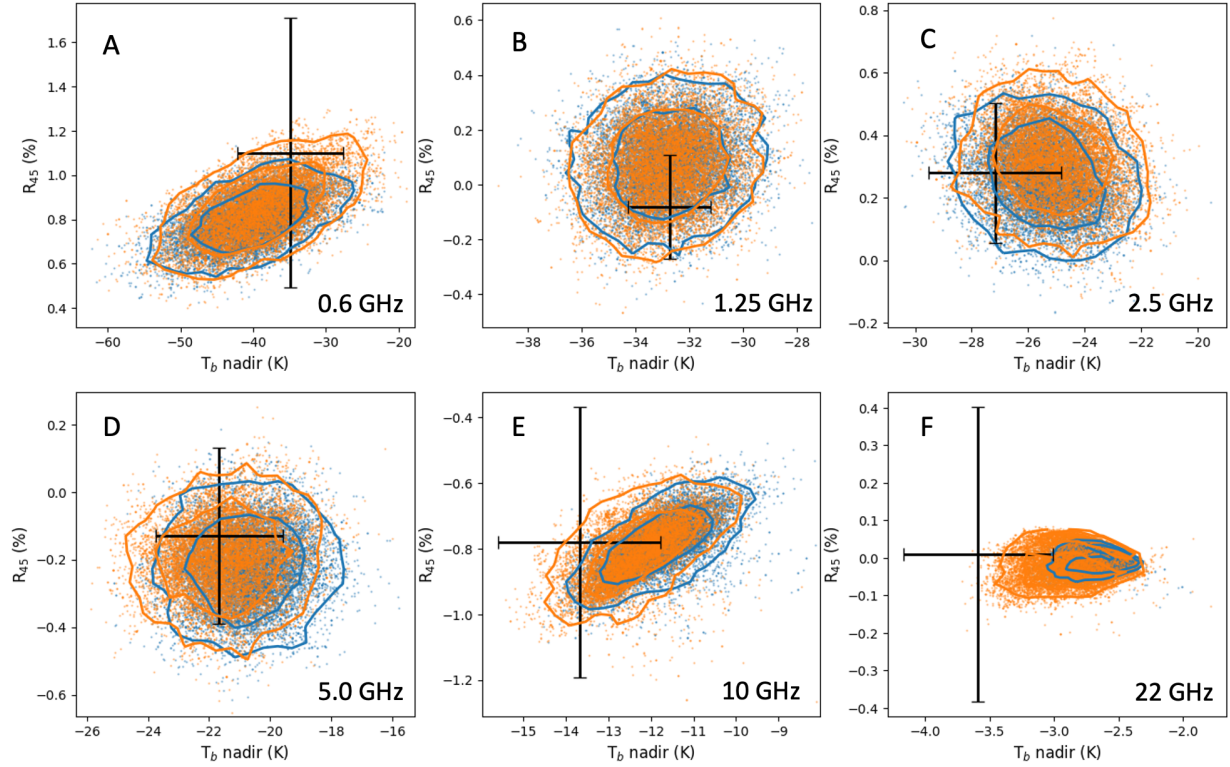

*Figure 8. MCMC samples states in both nadir brightness temperature and limb darkening space. Each dot represents a state in the MCMC chain, but with a 5-fold decrease in density. One and two sigma confidence intervals are drawn as two concentric contours. The blue contours represent the dry case (0.1 times solar water), while the orange contours represent the case with four times solar water. The Juno/MWR channel frequencies are indicated in the lower right corner of each panel. The black crosses signify the uncertainty of the equatorial anomaly which the MCMC inversion model fits.*

Finally, the degeneracy between the kinetic temperature and the ammonia concentration at three pressure levels (1 bar, 5 bar and 20 bar) is displayed as scatter plots in Figure 9. For the global-mean profile, a degeneracy between kinetic temperature and ammonia concentration is evident at all pressure levels. Specifically, at the 1-bar pressure level, a temperature increase of 5 K would produce a similar microwave observation as an increase in ammonia concentration by about 50 ppm (Figure 9A). The degeneracy becomes less significant for the contrast between the equatorial profile and the global mean. While the ammonia concentration can vary between 400 ~ 550 ppm (blue contours), Figure 9B shows that the temperature contrast at 5-bar (blue points) is readily determined at the precision of ±1 K with the equatorial temperature being colder than the global mean by about 7 K. The clustering of blue points near 275 K is driven by both the Juno/MWR observation and the stability requirement. First, $\chi^2$ fitting to the MWR observation is reduced if the temperature is comparatively low at 5 bar. Second, the atmosphere becomes unstable when the temperature is too low. As a result, the MCMC sampling finds the minimum temperature at 5-bar at which the atmosphere is marginally stable. At 20-bar, the uncertainty contours for both the equatorial profile and the global mean profile run parallel to each other. However, the ammonia concentration in the EZ is higher than the global mean by approximately 50 ppm.

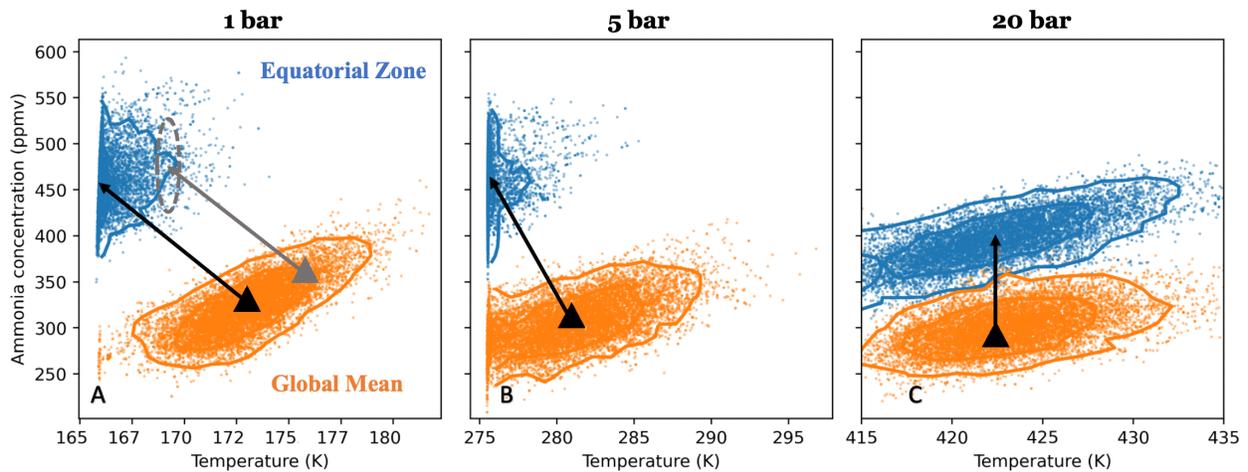

*Figure 9. Degeneracy between kinetic temperature and ammonia concentration at 1 bar, 5 bar and 20 bar pressure levels. The MCMC states sampled for the EZ are displayed in blue colors. The MCMC states sampled for the global mean are displayed in orange colors. From left to right, the pressure levels are at 1 bar, 5 bar and 20 bar. One-sigma and two-sigma uncertainties are drawn as concentric contours (ellipses). The meanings of arrows, dashed ellipses and triangles are explained in the main text.*

The high precision inference, 1 K out of 275 K, at 5-bar pressure level deserves a careful discussion. Recall that we have used two-stage differential fitting algorithm. In the first stage, we fit the global mean observation, taking into account the calibration uncertainty. In the second stage, we fit the equatorial anomaly without considering calibration uncertainty, leveraging the precision of the observation, which amounts to a fraction of a Kelvin. Consequently, the uncertainty estimation in the second stage pertains to *the difference from the global mean*, i.e.

$$\sigma(\delta T) = 1\ K \tag{6}$$

This explains why the spread of temperature uncertainty in the equatorial profile is much more constrained than in the global mean profile, as seen in Figures 9A and 9B. The triangle, arrow and circle in Figure 9A depict the relation between the global inversion (first stage) and the equatorial inversion (second stage). The triangle refers to where the truth is for the global profile, which can be anywhere inside the orange error ellipse. The difference between the equatorial profile and the global-mean profile is illustrated by the arrow. If the triangle shifts from black to grey, the origin of the arrow also shifts from black to grey but the length and direction of the arrow do not change. The uncertainty of the arrow is depicted by the dashed grey ellipse, which is much smaller than the orange ellipse.

Another observation is that the error ellipse at 20 bar pressure level (Figure 9C) does not behave in the same way as 1 bar (Figure 9A) and 5 bar (Figure 9B). This is because the uncertainty at the 20-bar pressure level is constrained by the precise limb darkening measurement at the 1.2-GHz channel, which is already a relative measurement. So, Figure 9C indicates that the EZ has more ammonia vapor than the global mean but their temperatures are similar at 20-bar ($422 \pm 4$ K). A 4 K uncertainty at 20-bar, when translated via the adiabatic relation, implies a 1.6 K uncertainty at the 1-bar pressure level. It is seen in Figure 7 that all temperature profiles gradually converge to an adiabat as the pressure approaches 100 bars of pressure. From there, we determine that the deep entropy of Jupiter is $\theta_{1bar} = 169 \pm 1.6$ K given all possibilities of ammonia concentration.

## 5 Estimation of the deep-water abundance based on the super-adiabatic temperature gradient

Since the Galileo Probe determined that, $T_{1bar} = 166 \pm 1$ K, we potentially observe an entropy difference between the shallow and deep levels:

$$T_{1bar} - \theta_{1bar} = -3.0 \pm 1.9\ K \tag{24}$$

This unstable temperature contrast can only be balanced by a vertical decrease of the mean molecular weight, which indicates the existence of water. In this section, we discuss the relationship of the water abundance to the potential temperature difference.

Contrary to what has sometimes been supposed, the moist adiabatic temperature profile is not a neutral stability profile, owing to the added effect of layered molecular weight, as given by the vertical gradient of the water mixing ratio. The temperature profile that is neutrally buoyant with respect to the vertical movement of air parcels, including latent heat, has been derived by Durran & Klemp (1982), confirming the earlier work of Lalas & Einaudi (1974). Richiardone & Giusti (2001) present an equivalent expression in which the vapor gradient is given in proportion to the undifferentiated mixing ratio, thereby providing explicit forms for the neutrally stable lapse rate (see their equations 7 and 9). Assuming convection adjusts a moist environment to neutrally buoyant stability, the integration of the modified lapse rate specifies the vertical temperature profile within a saturated region. As applied to terrestrial meteorology, where water is lighter than nitrogen-oxygen air, the neutrally stable lapse rate is only slightly steeper than the moist adiabat. Nevertheless, dropsonde statistics for precipitating cyclones in the eastern Pacific show better agreement with neutral stability (Richiardone & Manfrin, 2009) . But as applied here to Jupiter, where water is much heavier than hydrogen-helium air, the neutrally stable lapse rate is steeper, or super-adiabatic, and by an amount depending upon the water abundance.

Figure 10 shows the vertical profiles of neutrally buoyant potential temperature within Jupiter's cloud layer, calculated by vertical integration of the Richiardone-Giusti lapse rate equations, for the labeled abundances of water, given in proportion to the solar oxygen to hydrogen ratio. A vertical potential temperature difference of $|\Delta\theta| = 3\ K$ corresponds to a water abundance of about 5 times solar. Note that most of the vertical variation occurs within a thin layer between the 6.5 and 4 bar pressure levels, surmounted by an approximately adiabatic lapse rate aloft. However, these results, based on the neutrally stable lapse-rate equations, does not necessarily assume a fully saturated environment, and neglect the effects of moist entrainment.

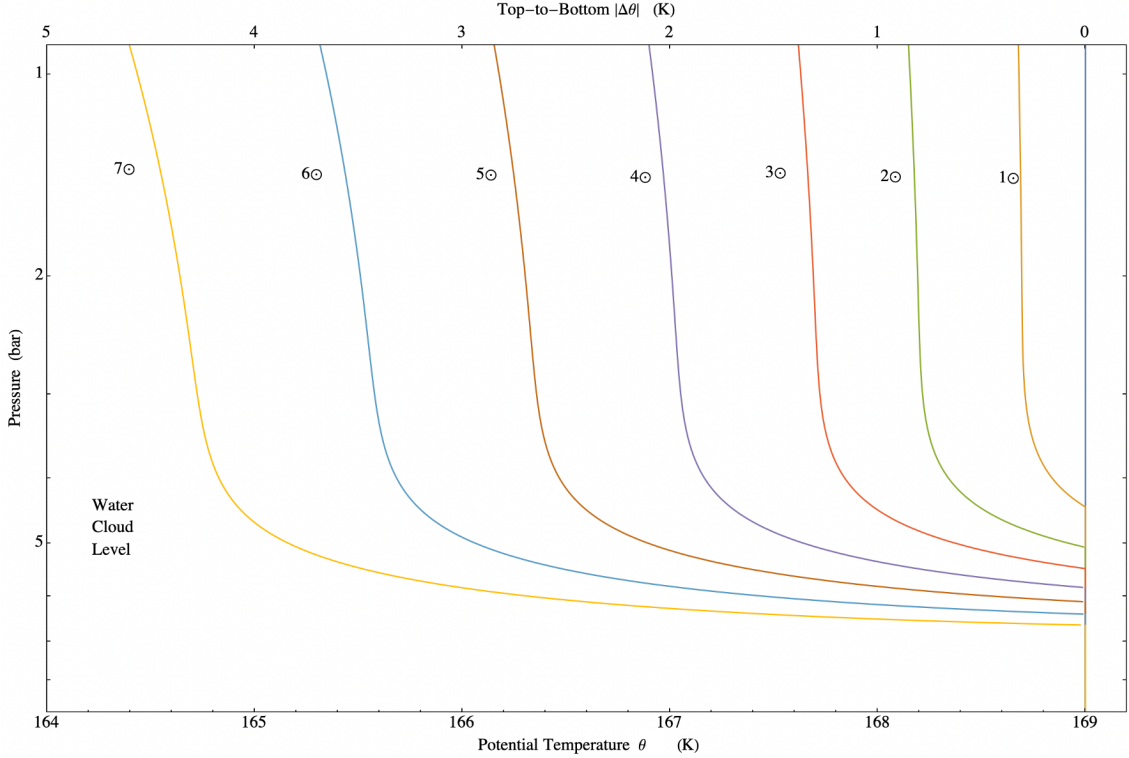

*Figure 10. Potential temperature profiles of a neutrally stratified saturated atmosphere, as calculated by integration of the Richiardone-Giusti lapse rate equations, for the indicated choices of the deep water abundance.*

A more realistic estimation, taking the updrafts, downdrafts and mixing into account, can be obtained by 3D numerical simulation of moist convection. Adapting the SNAP (Simulating Non-hydrostatic Atmospheres on Planets) model, we have conducted a series of cloud-resolving simulations with different deep water abundance by expanding the published 2D version (Li & Chen, 2019) to 3D (Ge et al., 2023). We use the same forcing and evolve the model to a steady state. Then, we gather the domain-averaged temperature profile and estimate its mean and standard deviation over time. We conducted four experiments at 0.1x, 0.3x, 1x and 3x solar water abundances. The resulting relationship derived from the numerical experiment and uncertainties are shown as the orange curve and shaded are of Figure 11 and is very similar to the Richiardone-Giusti neutral-lapse-rate relation plotted as the blue curve. A least-square fit (correlation coefficient, $R = 0.99$) to the simulation results yields the simple linear relationship:

$$\Delta\theta = (0.60 \pm 0.14)K \times \left(\frac{x}{x_\odot}\right) + (0.04 \pm 0.02)K. \tag{25}$$

Since the Juno/MWR-inferred $\Delta\theta = 3.0 \pm 1.9\ K$, given by the statistical estimates of the slope and the intercept, the distribution function of $\left(\frac{x}{x_\odot}\right)$ is a quotient distribution by two noncentral normal distributions. Let two normal variables be $X = N(\mu_X, \sigma_X^2)$ and $Y = N(\mu_Y, \sigma_Y^2)$, the ratio $Z = X/Y$ can be approximated as a normal distribution with mean $\mu_Z = \mu_X/\mu_Y$ and variance $\sigma_Z^2 = \left(\frac{\mu_X}{\mu_Y}\right)^2 \left(\frac{\sigma_X^2}{\mu_X^2} + \frac{\sigma_Y^2}{\mu_Y^2}\right)$ (Díaz-Francés & Rubio, 2013). Using the above equations, the statistical range of $\left(\frac{x}{x_\odot}\right)$ is:

$$\left(\frac{x}{x_\odot}\right) = 4.9 \pm 3.4. \qquad (26)$$

Consequently, we estimate a water enrichment of approximately $1.5 \sim 8.3 \times$ solar at Jupiter's equator. This new result has a similar upper bound as estimated by our previous work assuming a moist adiabat (Li et al., 2020), but rules out a solar or subsolar abundance of water. It does, however, allow for a C/O ratio that is solar, given the similar enrichment of carbon. A solar C/O value in the Jovian atmosphere would be consistent with the solar-composition icy planetesimal model of (Owen & Encrenaz, 2003).

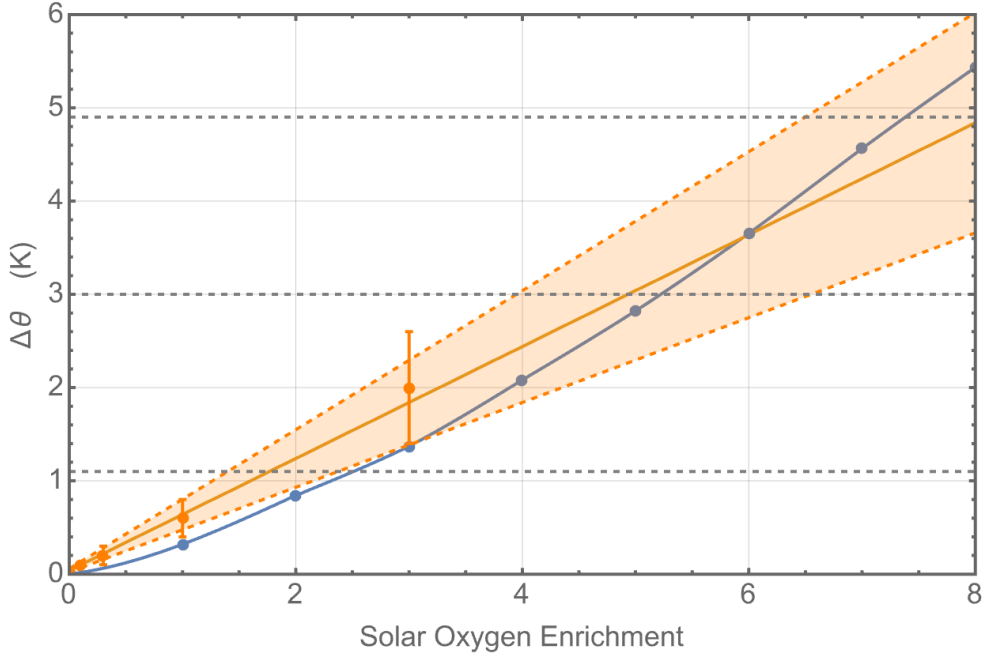

Figure 11. *The relation between the potential temperature difference and the oxygen enrichment. Four dots with error bars are statistical results from three-dimensional numerical simulations at four different deep water abundance. The orange line and shaded area show the best-fit to*

*numerical simulations and the estimated uncertainty range (equation 8). Blue dots are theoretical calculations based on integrating a neutral density lapse rate (translation of Figure 10). Three horizontal lines are $\Delta\theta = 1.1\ K, 3\ K, 4.9\ K$ respectively.*

## 6 Conclusion

There are three key findings from this study. First, we determined that the deep entropy of Jupiter's atmosphere is $\theta_{1bar} = 169 \pm 1.6$ K. This result comes from a careful study of the brightness temperature and limb darkening observations made by the Juno/MWR instrument from the first 12 perijove passes of Jupiter. The precise limb darkening measured by the 1.2 GHz channel plays a crucial rule in narrowing the uncertainty.

Second, we detected a super-adiabatic temperature gradient near Jupiter's EZ. A quantitative estimate is that $T_{1bar} - \theta_{1bar} = -3.0 \pm 1.9\ K$. This result relies on both Juno/MWR's probe of the deep atmosphere and the Galileo Probe's in-situ measurement of the temperature at the 1-bar pressure level.

Third, the super-adiabatic temperature gradient highlighted in our second finding suggests a water abundance exceeding solar values. Using 3D numerical simulations, we have determined an empirical relation $\Delta\theta \approx (0.60 \pm 0.14)\ K \times \left(\frac{x}{x_\odot}\right)$. This empirical relation agrees with the one-dimensional result obtained by vertical integration of the Richiardone-Giusti lapse rate equations. Using the empirical relation and the super-adiabatic temperature gradient leads to an estimated deep-water abundance ranging from 1.5 to 8.3 times the solar value.

## 7 Caveats of this study, discussion and looking forward

Juno's MWR instrument records the radio emissions from various depths within Jupiter's atmosphere. To interpret these MWR data in terms of physical attributes, such as its composition and temperature, the contextual (or ambient) information, utilization of statistical methods and numerical simulations are all as vital as the MWR measurements themselves.

We have employed temperature and ammonia variability as the prior constraints, used the Galileo Probe results as a reference point, constructed atmospheric profiles using Gaussian Process, and utilized meteorological theory and numerical simulations to establish a new constraint on the water abundance. Should any of these factors prove unreliable, the conclusions presented in this article may need revision based on updated ambient information. In this context, we offer forward-looking discussions and emphasize the cautions and caveats tied to this research.

**Discussion #1. The Galileo Probe observation**

The Galileo Probe provides the sole in-situ measurement of a giant planet's atmosphere to date. While it functioned up to a pressure of 20 bars and assessed the atmospheric conditions at a single, anomalous location on Jupiter – a 5-$\mu m$ hot spot – its data remains invaluable. It offers a crucial reference point for all indirect analyses of Jupiter's atmosphere based on remote observations. Understanding the origin of the hot spot, the underlying physical mechanism, and how representative its findings are of Jupiter's entire atmosphere is of utmost importance.

From the study of the Juno MWR data, we find that the Galileo Probe site, a 5-$\mu m$ hot spot, is *atypical* in the following ways:

(1) The composition (particularly ammonia vapor) measured by the Galileo Probe is different from either the global mean or the EZ. The global-mean deep ammonia concentration is revised to $310 \pm 20$ ppm based in this study, which is less than the lower limit of the Galileo Probe in-situ observation (350 ppm). The discrepancy could be due to an inverted ammonia gradient near the equator, which is confirmed by multiple inversion study of the Juno observation (Li et al., 2020; Moeckel et al., 2023).

(2) The composition profile measured by the Galileo Probe is different from either the global mean or the EZ. The vertical profile of ammonia vapor measured by the Galileo Probe is similar to what one would expect from an equilibrium condensation model but displaced to a higher pressure. Li et al., (2018) constructed a parameterized "adiabatic stretching" model in which the atmosphere is stretched adiabatically in the vertical by a factor of $X$. Using an $X = 4$ stretching – an air parcel is displaced from 1 bar to 4 bars – they were able to fit the profile of both NH$_3$ and H$_2$S (Figure 3 of Li et al., 2018). However, the stretching model does not appear to be globally applicable. Neither the global mean nor the EZ shows a significant stretching of air column. It is likely that significant stretching (vertical displacement) only exists at the 5-$\mu m$ hot spot (Showman & Dowling, 2000).

(3) The temperature profile recorded by the Galileo Probe suggests the level of non-adiabaticity in Jupiter's atmosphere, but it may not represent the global average or the average of the EZ as a whole. The Galileo Probe ASI returned two reliable pieces of information regarding Jupiter's atmosphere. The first is the temperature at 6 °N and at 1-bar pressure level, $T_{1bar} = 166.1 \pm 0.8\ K$. A thorough re-analysis of the Voyager radio occultation data yielded roughly the same temperature at the equator (Gupta et al., 2022; Lindal et al., 1981). However, the temperature at 12 °S is 4 K higher (Gupta et al., 2022). So, indeed, the EZ could be a cold place on the planet compared to the neighboring latitudes. The second observation concerns the variability in the atmospheric temperature profile between 1 bar and 20 bars as the probe descended. This finding has frequently been overlooked in publications due to its complex nature and challenging interpretation. In this article, we highlight the significance of this observation, emphasizing that the 0.1~0.2 K/km non-adiabaticity detected by the Galileo Probe should not be dismissed. In particular, we found that globally averaged temperature profile is sub-adiabatic at the level of 0.1 K/km between 1 to 100 bars and the EZ is super-adiabatic across the water condensation level.

(4) The temperature measurement by the Galileo Probe at 1-bar pressure level cannot be used as a value for the deep entropy. This study of the Juno MWR observations reveals that the EZ of Jupiter may exhibit a stable super-adiabatic temperature gradient across the water condensation level. If the Galileo Probe profile can be interpreted by the stretching model discussed in point (2), stretching the background shallow atmosphere by a factor of 4 in pressure leads to a colder temperature at depth compared to the unperturbed atmosphere. However, the colder temperature does not mean that the air in the 5-$\mu m$ hot spot is denser because the 5-$\mu m$ hot spot is a dry spot and the background atmosphere is water-rich and thus high in density.

**Discussion #2. The Deep-Water Abundance**

We infer the deep-water abundance based on the density contrast between the shallow and deep atmosphere. Our nominal estimate of the 4.9 times solar oxygen enrichment is comparable to the well-established value for carbon (Niemann et al., 1998; Owen & Encrenaz, 2003; Wong et al., 2004, 2008). Our inference hinges on these four conditions: (1) The atmosphere of Jupiter is stably stratified. (2) The deep entropy of Jupiter is very nearly uniform across all latitudes. (3) The temperature at the 1-bar pressure level is known from observations other than from the Juno MWR. (4) The super-adiabatic temperature gradient (if any) linearly scales with the deep water abundance. The first condition is met by fluid dynamics. The second condition is a plausible assumption, consistent with deep convective mixing in the absence of a rigid lower boundary. The third condition is met by the Galileo Probe observation and the Voyager radio occultation experiment near the equator. And the fourth condition is supported by established theory and numerical simulations over a limited range of assumed water abundances.

Although the inferred 1.5 ~ 8.3 enrichment of O/H ratio at Jupiter's equatorial region is derived under various assumptions, the presented method offers a new way of detecting water on Jupiter through the measurement of the potential temperature difference between the shallow and deep layers.

A supersolar value of water has implications for the nature of the deeper interior. At kilobar pressures, Cavalié et al. (2023) derived the water abundance from modeling the kinetics of $CO$-$CH_4$ conversion. Their results are consistent with supersolar water only if the vigor of vertical mixing in the kilobar region is very sluggish, as might be obtained by a deep radiative zone (Guillot et al., 1994) or other mechanism to inhibit convection. Interior models tied to Juno gravity data severely limit the abundance of heavy elements in the deep envelope (hundreds of kiolobars to megabars), such that twice solar oxygen or larger is not permitted (Miguel et al., 2022; Militzer et al., 2022). Reconciliation of the results might require an outer envelope with an oxygen (water) abundance different from that in the deeper interior (Helled et al., 2022).

**Discussion #3. Spectral inversion method**

In this work, we develop a novel spectral inversion method to infer the atmospheric temperature structure from microwave observations. This method forms the cornerstone for deducing the deep water abundance in the study. We first demonstrate that Juno MWR's brightness temperature and limb darkening measurement contains information about the kinetic temperature structure and the constraint on temperature is stronger at long-wavelength channels because (1) the variability of ammonia decreases as one goes deeper into the atmosphere and (2) a small non-adiabatic temperature gradient is amplified by integrating over a large vertical distance.

In order to break the degeneracy between the ammonia profile and the temperature profile, we develop ways to leverage the precision of the measurement by invoking a two-stage fitting, one for the global average, and another for the equatorial anomaly. To regularize the variability of the ammonia profile, we utilize the Gaussian Process to force a vertical correlation between ammonia concentrations. To allow a wider possibility of the temperature profiles, we randomly sample the atmosphere at evenly spaced levels and construct a smooth interpolated profile. To exclude impossible solutions of the temperature structure, we make corrections to the sampled atmospheric profile wherever the atmosphere is unstable to convection. The flexibility, regularity and the robustness of the current method supersedes the previous publications on the Juno MWR retrievals (Li et al., 2017, 2020).

The spectral inversion method detailed in this article employs advanced statistical techniques. These techniques warrant further examination by statisticians and mathematicians, optimization by computer scientists, and validation through alternative measurement methods. Harnessing the MWR instrument's accuracy and precision using statistical approaches still remains a largely untapped area.

**Discussion #4. Origin of super-adiabaticity**

The two-dimensional cloud-resolving simulation of Jovian weather by Li & Chen (2019) demonstrates that a super-adiabatic temperature gradient naturally develops in the presence of water when the atmosphere is made of hydrogen and is heated from bottom and cooled from top.

This type of simulation best represents the equatorial region of Jupiter due to the absence of the Coriolis force.

Many similar theoretical works have been published suggesting the existence of a stable super-adiabatic layer in hydrogen atmospheres. To name a few, Guillot (1995) suspected that, for a hydrogen atmosphere loaded with heavy condensable species, there is a threshold above which the vertical gradient of molecular weight due to condensation can stabilize the layer against moist convection. Li & Ingersoll (2015) pointed out that a stable super-adiabatic layer is crucial for regulating the periodic eruption of Saturn's giant storms. and thus deep water abundance on Saturn must exceed a value of 10 times solar. Friedson & Gonzales (2017) and Markham & Stevenson (2021) considered the inhibition of thermal convection in condensate-rich planets such as Uranus and Neptune and the implication for their thermal evolution history.

Although it has been theorized for decades, this study provides the first detection of a stable super-adiabatic temperature gradient in a giant planet via the study of microwave spectra. The origin of the super-adiabaticity is due to the contrast in molecular weight between the light ambient atmosphere made of hydrogen and helium and the heavy condensable volatiles made of water and ammonia. The super-adiabatic temperature layer may be ubiquitous for all giant planets. The analysis of Juno MWR data suggests that the EZ is most likely to exhibit a super-adiabatic temperature gradient where higher latitudes may have other unique dynamics that suppresses the development of super-adiabatic temperature gradient.

**Discussion #5. Joint multi-instrument analysis**

Our analysis benefits from combining Juno and Galileo probe measurements. The accuracy of Galileo anchors the uncertainty in the temperature at low pressure, where ammonia is highly variable, and the precision of Juno's MWR constrains the temperature of the deep atmosphere where there is less variability. Additional complementary measurements are possible such as space- and ground-based infrared observations, radio occultations, among others. This article presents the framework for a combined analysis of the MWR data at depth and the Galileo entry probe data at shallower levels. A joint analysis of the MWR data with the infrared observations could potentially reveal the temperature structure at the 1-bar pressure level for a wider range of latitudes. This approach will be the subject of future work.

**Acknowledgement**

CL and SA are supported by the NASA Juno Program, under NASA Contract NNM06AA75C from the Marshall Space Flight Center, through subcontract 699056KC and Q99063JAR to the University of Michigan from the Southwest Research Institute. GO and SL are supported by the Juno mission under the NASA Contract 80NM0018D0004. Y.M. acknowledges funding from the European Research Council (ERC) under the European Union's Horizon 2020 research and innovation programme (grant agreement no. 101088557, N-GINE). Finally, we thank Michael Janssen for designing such a beautiful, innovative and exquisitely sensitive instrument as the Juno Microwave Radiometer, and for his leadership of the MWR team 2003 through 2017.

**References**

Arakawa, A., & Schubert, W. H. (1974). Interaction of a cumulus cloud ensemble with the large-scale environment, Part I. *Journal of the Atmospheric Sciences*, *31*(3), 674–701.

Asplund, M., Grevesse, N., Sauval, A. J., & Scott, P. (2009). The Chemical Composition of the Sun. *Annual Review of Astronomy and Astrophysics*, *47*(1), 481–522. https://doi.org/10.1146/annurev.astro.46.060407.145222

Bellotti, A., Steffes, P. G., & Chinsomboom, G. (2016). Laboratory measurements of the 5–20 cm wavelength opacity of ammonia, water vapor, and methane under simulated conditions for the deep jovian atmosphere. *Icarus*, *280*, 255–267.

Betts, A. K., & Miller, M. J. (1993). The Betts-Miller scheme. In *The representation of cumulus convection in numerical models* (pp. 107–121). Springer.

Cavalié, T., Lunine, J., & Mousis, O. (2023). A subsolar oxygen abundance or a radiative region deep in Jupiter revealed by thermochemical modelling. *Nature Astronomy*, 1–6.

Conrath, B. J., & Gierasch, P. J. (1984). Global variation of the para hydrogen fraction in Jupiter's atmosphere and implications for dynamics on the outer planets. *Icarus*, *57*(2), 184–204.

Díaz-Francés, E., & Rubio, F. J. (2013). On the existence of a normal approximation to the distribution of the ratio of two independent normal random variables. *Statistical Papers*, *54*, 309–323.

Duong, D., Steffes, P. G., & Noorizadeh, S. (2014). The microwave properties of the jovian clouds: A new model for the complex dielectric constant of aqueous ammonia. *Icarus*, *229*, 121–130.

Durran, D. R., & Klemp, J. B. (1982). On the effects of moisture on the Brunt-Väisälä frequency. *Journal of Atmospheric Sciences*, *39*(10), 2152–2158.

Emanuel, K. A. (1994). *Atmospheric convection*. New York: Oxford University Press.

Fletcher, L. N., Orton, G. S., Mousis, O., Yanamandra-Fisher, P., Parrish, P. D., Irwin, P. G. J., et al. (2010). Thermal structure and composition of Jupiter's Great Red Spot from high-resolution thermal imaging. *Icarus*, *208*(1), 306–328.

Fletcher, L. N., Greathouse, T., Orton, G., Sinclair, J., Giles, R., Irwin, P., & Encrenaz, T. (2016). Mid-infrared mapping of Jupiter's temperatures, aerosol opacity and chemical distributions with IRTF/TEXES. *Icarus*, *278*, 128–161.

Folkner, W. M., Woo, R., & Nandi, S. (1998). Ammonia abundance in Jupiter's atmosphere derived from the attenuation of the Galileo probe's radio signal. *Journal of Geophysical Research: Planets*, *103*(E10), 22847–22855.

Friedson, A. J., & Gonzales, E. J. (2017). Inhibition of ordinary and diffusive convection in the water condensation zone of the ice giants and implications for their thermal evolution. *Icarus*, *297*, 160–178.

Galanti, E., Kaspi, Y., Duer, K., Fletcher, L., Ingersoll, A. P., Li, C., et al. (2021). Constraints on the latitudinal profile of Jupiter's deep jets. *Geophysical Research Letters*, *48*(9), e2021GL092912.

Ge, H., Li, C., & Xi, Z. (2023). A stable and superadiabatic weather layer in Jupiter induced by moist convection. *Submitted to Nature Astronomy*.

Gierasch, P. J., Conrath, B. J., & Magalhães, J. A. (1986). Zonal mean properties of Jupiter's upper troposphere from Voyager infrared observations. *Icarus*, *67*(3), 456–483.

Gierasch, P. J., Conrath, B. J., & Read, P. L. (2004). Nonconservation of Ertel potential vorticity in hydrogen atmospheres. *Journal of the Atmospheric Sciences*, *61*(15), 1953–1965.

Goodman, J., & Weare, J. (2010). Ensemble samplers with affine invariance. *Communications in Applied Mathematics and Computational Science*, *5*(1), 65–80.

Guillot, T., Gautier, D., Chabrier, G., & Mosser, B. (1994). Are the Giant Planets Fully Convective? *Icarus*, *112*(2), 337–353. https://doi.org/10.1006/icar.1994.1188

Guillot, T., Chabrier, G., Morel, P., & Gautier, D. (1994). Nonadiabatic models of Jupiter and Saturn. *Icarus*, *112*(2), 354–367.

Guillot, Tristan. (1995). Condensation of Methane, Ammonia, and Water and the Inhibition of Convection in Giant Planets. *Science*, *269*(5231), 1697–1699. https://doi.org/10.1126/science.7569896

Gupta, P., Atreya, S. K., Steffes, P. G., Fletcher, L. N., Guillot, T., Allison, M. D., et al. (2022). Jupiter's Temperature Structure: A Reassessment of the Voyager Radio Occultation Measurements. *arXiv Preprint arXiv:2205.12926*.

Hanley, T. R., Steffes, P. G., & Karpowicz, B. M. (2009). A new model of the hydrogen and helium-broadened microwave opacity of ammonia based on extensive laboratory measurements. *Icarus*, *202*(1), 316–335.

Helled, R., Stevenson, D. J., Lunine, J. I., Bolton, S. J., Nettelmann, N., Atreya, S., et al. (2022). Revelations on Jupiter's formation, evolution and interior: Challenges from Juno results. *Icarus*, *378*, 114937.

Holton, J. R. (1973). An introduction to dynamic meteorology. *American Journal of Physics*, *41*(5), 752–754.

Janssen, M. A., Hofstadter, M. D., Gulkis, S., Ingersoll, A. P., Allison, M., Bolton, S. J., et al. (2005). Microwave remote sensing of Jupiter's atmosphere from an orbiting spacecraft. *Icarus*, *173*(2), 447–453.

Janssen, M. A., Oswald, J. E., Brown, S. T., Gulkis, S., Levin, S. M., Bolton, S. J., et al. (2017). MWR: Microwave radiometer for the Juno mission to Jupiter. *Space Science Reviews*, *213*(1–4), 139–185.

Lalas, D. P., & Einaudi, F. (1974). On the correct use of the wet adiabatic lapse rate in stability criteria of a saturated atmosphere. *Journal of Applied Meteorology and Climatology*, *13*(3), 318–324.

Leconte, J., Selsis, F., Hersant, F., & Guillot, T. (2017). Condensation-inhibited convection in hydrogen-rich atmospheres-Stability against double-diffusive processes and thermal profiles for Jupiter, Saturn, Uranus, and Neptune. *Astronomy & Astrophysics*, *598*, A98.

Li, C., & Chen, X. (2019). Simulating Nonhydrostatic Atmospheres on Planets (SNAP): Formulation, Validation, and Application to the Jovian Atmosphere. *The Astrophysical Journal Supplement Series*, *240*(2), 37.

Li, C., & Ingersoll, A. P. (2015). Moist convection in hydrogen atmospheres and the frequency of Saturn's giant storms. *Nature Geoscience*, *8*(5), 398–403. https://doi.org/10.1038/ngeo2405

Li, C., Ingersoll, A., Janssen, M., Levin, S., Bolton, S., Adumitroaie, V., et al. (2017). The distribution of ammonia on Jupiter from a preliminary inversion of Juno microwave radiometer data. *Geophysical Research Letters*, *44*(11), 5317–5325.

Li, C., Le, T., Zhang, X., & Yung, Y. L. (2018). A high-performance atmospheric radiation package: With applications to the radiative energy budgets of giant planets. *Journal of Quantitative Spectroscopy and Radiative Transfer*, *217*, 353–362.

Li, C., Ingersoll, A. P., & Oyafuso, F. (2018). Moist Adiabats with Multiple Condensing Species: A New Theory with Application to Giant-Planet Atmospheres. *Journal of the Atmospheric Sciences*, *75*(4), 1063–1072. https://doi.org/10.1175/JAS-D-17-0257.1

Li, C., Ingersoll, A., Bolton, S., Levin, S., Janssen, M., Atreya, S., et al. (2020). The water abundance in Jupiter's equatorial zone. *Nature Astronomy*, *4*, 609–616.

Li, C., de Pater, I., Moeckel, C., Sault, R. J., Butler, B., deBoer, D., & Zhang, Z. (2023). Long-lasting, deep effect of Saturn's Giant Storms. *Science Advances*, *9*(32), eadg9419.

Lindal, G. F., Wood, G., Levy, G., Anderson, J., Sweetnam, D., Hotz, H., et al. (1981). The atmosphere of Jupiter: An analysis of the Voyager radio occultation measurements. *Journal of Geophysical Research: Space Physics*, *86*(A10), 8721–8727.

Magalhães, J. A., Seiff, A., & Young, R. E. (2002). The stratification of Jupiter's troposphere at the Galileo probe entry site. *Icarus*, *158*(2), 410–433.

Manabe, S., & Wetherald, R. T. (1975). The effects of doubling the CO2 concentration on the climate of a general circulation model. *Journal of the Atmospheric Sciences*, *32*(1), 3–15.

Markham, S., & Stevenson, D. (2021). Constraining the effect of convective inhibition on the thermal evolution of Uranus and Neptune. *The Planetary Science Journal*, *2*(4), 146.

Miguel, Y., Bazot, M., Guillot, T., Howard, S., Galanti, E., Kaspi, Y., et al. (2022). Jupiter's inhomogeneous envelope. *arXiv Preprint arXiv:2203.01866*.

Militzer, B., Hubbard, W. B., Wahl, S., Lunine, J. I., Galanti, E., Kaspi, Y., et al. (2022). Juno spacecraft measurements of Jupiter's gravity imply a dilute core. *The Planetary Science Journal*, *3*(8), 185.

Moeckel, C., de Pater, I., & DeBoer, D. (2023). Ammonia Abundance Derived from Juno MWR and VLA Observations of Jupiter. *The Planetary Science Journal*, *4*(2), 25.

Niemann, H., Atreya, S., Carignan, G., Donahue, T., Haberman, J., Harpold, D., et al. (1998). The composition of the Jovian atmosphere as determined by the Galileo probe mass spectrometer. *Journal of Geophysical Research: Planets*, *103*(E10), 22831–22845.

Owen, T., & Encrenaz, T. (2003). Element abundances and isotope ratios in the giant planets and Titan. *Space Science Reviews*, *106*(1–4), 121–138.

Oyafuso, F., Levin, S., Orton, G., Brown, S. T., Adumitroaie, V., Janssen, M., et al. (2020a). Angular Dependence and Spatial Distribution of Jupiter's Centimeter-Wave Thermal Emission From Juno's Microwave Radiometer. *Earth and Space Science*, *7*(11), e2020EA001254.

Oyafuso, F., Levin, S., Orton, G., Brown, S. T., Adumitroaie, V., Janssen, M., et al. (2020b). Angular Dependence and Spatial Distribution of Jupiter's Centimeter-Wave Thermal Emission From Juno's Microwave Radiometer. *Earth and Space Science*, *7*(11), e2020EA001254.

de Pater, I., Dunn, D., Romani, P., & Zahnle, K. (2001). Reconciling Galileo probe data and ground-based radio observations of ammonia on Jupiter. *Icarus*, *149*(1), 66–78.

de Pater, I., Sault, R., Butler, B., DeBoer, D., & Wong, M. H. (2016). Peering through Jupiter's clouds with radio spectral imaging. *Science*, *352*(6290), 1198–1201.

de Pater, I., Sault, R. J., Wong, M. H., Fletcher, L. N., DeBoer, D., & Butler, B. (2019). Jupiter's ammonia distribution derived from VLA maps at 3–37 GHz. *Icarus*, *322*, 168–191.

Richiardone, R., & Giusti, F. (2001). On the stability criterion in a saturated atmosphere. *Journal of the Atmospheric Sciences*, *58*(14), 2013–2017.

Richiardone, R., & Manfrin, M. (2009). Neutral saturated lapse rate: an experimental check from CALJET-1998 and PACJET-2001. *Monthly Weather Review*, *137*(12), 4382–4385.

Rodgers, C. D. (2000). *Inverse methods for atmospheric sounding: theory and practice* (Vol. 2). World scientific.

Sault, R. J., Engel, C., & de Pater, I. (2004). Longitude-resolved imaging of Jupiter at λ= 2 cm. *Icarus*, *168*(2), 336–343.

Schneider, T., & Liu, J. (2009). Formation of jets and equatorial superrotation on Jupiter. *Journal of the Atmospheric Sciences*, *66*(3), 579–601.

Seiff, A., Kirk, D. B., Knight, T. C., Young, R. E., Mihalov, J. D., Young, L. A., et al. (1998). Thermal structure of Jupiter's atmosphere near the edge of a 5-μm hot spot in the north equatorial belt. *Journal of Geophysical Research: Planets*, *31*(E10), 22857–22889.

Showman, A. P., & Dowling, T. E. (2000). Nonlinear simulations of Jupiter's 5-micron hot spots. *Science*, *289*(5485), 1737–1740.

Wong, M. H., Mahaffy, P. R., Atreya, S. K., Niemann, H. B., & Owen, T. C. (2004). Updated Galileo probe mass spectrometer measurements of carbon, oxygen, nitrogen, and sulfur on Jupiter. *Icarus*, *171*(1), 153–170.

Wong, M. H., Lunine, J. I., Atreya, S. K., Johnson, T., Mahaffy, P. R., Owen, T. C., & Encrenaz, T. (2008). Oxygen and other volatiles in the giant planets and their satellites. *Reviews in Mineralogy and Geochemistry*, *68*(1), 219–246.

Xu, K., & Emanuel, K. A. (1989). Is the tropical atmosphere conditionally unstable? *Monthly Weather Review*, *117*(7), 1471–1479.

Young, R. E., Smith, M. A., & Sobeck, C. K. (1996). Galileo probe: In situ observations of Jupiter's atmosphere. *Science*, *272*(5263), 837–838.